\documentclass[useAMS,usenatbib,usegraphicx]{mn2e}

\title[The age of NGC\,2547]{The lithium depletion boundary and the age of
NGC\,2547}
\author[Oliveira et al.]{J.M.
Oliveira$^{1}$\thanks{E-mail:joana@astro.keele.ac.uk},
R.D. Jeffries$^{1}$, C.R. Devey$^{1}$\thanks{Nuffield Foundation
Undergraduate Research Bursar (NUF-URB00)}, D. Barrado y
Navascu\'{e}s$^{2}$, \newauthor T. Naylor$^{3}$,
J.R. Stauffer$^{4}$  and E.J. Totten$^{1}$
\\
$^{1}$School of Chemistry \& Physics, Keele University, Keele, Staffordshire, ST5 5BG,
UK\\
$^{2}$ LAEFF-INTA, P.O. Box 50727, E-28080 Madrid, Spain\\
$^{3}$ School of Physics, University of Exeter, Stocker Road, Exeter,
EX4 4QL, UK\\
$^{4}$ SIRTF Science Center, California Institute of Technology, MS
314-6, Pasadena, CA 91125, USA}
\begin{document}

\date{Accepted 2003. Received 2003; in original form 2002}

\pagerange{\pageref{firstpage}--\pageref{lastpage}} \pubyear{2003}

\maketitle

\label{firstpage}

\begin{abstract}
We present the results of a photometric and spectroscopic survey of
cool M dwarf candidates in the young open cluster NGC\,2547. Using the
2dF fiber spectrograph, we have searched for the luminosity at which
lithium remains unburned in an attempt to constrain the cluster age.
The lack of a population of individual lithium-rich objects towards the faint end of our sample places a very strong lower limit to the cluster age of
35\,Myr. However, the detection of lithium in the averaged spectra of our
faintest targets suggests that the lithium depletion boundary lies at
$9.5<M_{I}<10.0$ and that the cluster age is $<54$\,Myr. The age of NGC\,2547 judged from fitting isochrones to low-mass pre-main-sequence stars
in colour-magnitude diagrams is 20$-$35\,Myr using the same
evolutionary models. The sense and size of the discrepancy in age
determined by these two techniques is similar to that found in another
young cluster, IC\,2391, and in the low-mass pre main-sequence binary
system, GJ\,871.1AB. We suggest that the inclusion of rotation or
dynamo-generated magnetic fields in the evolutionary models could
reconcile the two age determinations, but only at the expense of
increasing the cluster ages beyond that currently indicated by the lithium
depletion. Alternatively, some mechanism is required that
increases the rate of lithium depletion in young, very low-mass fully convective stars.
\end{abstract}


\section{Introduction}
Lithium can be a very short-lived element in the atmospheres of low mass
pre-main-sequence (PMS) stars, burning swiftly in $p,\alpha$ reactions once core temperatures approach $3\times 10^{6}$\,K \citep{ushomirsky98}. Because PMS stars less massive than 0.35\,M$_{\odot}$ are always fully convective, the surface abundance of Li is also rapidly depleted when the core reaches this Li-burning temperature. The time taken for this to happen is sensitively dependent on mass and to a lesser extent on adopted equations of state and atmospheric boundary conditions. As a result, if one observes a cluster of co-eval stars with a range of properties, the mass, and hence luminosity and temperature, at which Li is observed to be depleted from its initial value offers a potentially precise determination of the cluster age. It has been claimed that this technique is less subject to systematic uncertainties in stellar models than cluster ages derived from the main-sequence turn-off or fitting low-mass, PMS isochrones (e.g.\ \citealt{bildsten97}).

Several attempts have been made to define the lithium depletion boundary
(hereafter LDB) in the Pleiades, the $\alpha$\,Per cluster and IC\,2391.
\citet*{stauffer98} derive an LDB age of $125\pm8$\,Myr for the Pleiades,
considerably older than the nuclear turn-off age of 78\,Myr derived by
\citet{mermilliod81} using models with no convective core overshoot, but younger
than the 150\,Myr obtained by \citet{mazzei88} using models with very strong overshoot. Similarly, the LDB ages of $90\pm10$\,Myr and $53\pm5$\,Myr derived for the $\alpha$\,Per \citep{stauffer99} and IC\,2391 clusters \citep*{navascues99} are older than their zero-overshoot nuclear turn-off ages of 51\,Myr and 36\,Myr respectively. The LDB ages offer a powerful argument for a modest amount of convective core overshoot and agree reasonably well with nuclear turn-off ages from such models \citep{ventura98}. Measuring the LDB in clusters of different ages and different turn-off masses might reveal whether the amount of convective overshoot is mass-dependent.

Finding clusters which are amenable to such analysis is not easy. The LDB position is difficult to measure because of the intrinsic faintness of the relevant PMS
stars. Neither is the technique sensitive to ages less than 10\,Myr or older
than about 300\,Myr. In the former case, lithium remains unburned in stars of
all masses, whereas in the latter the gradient of the relationship between the
luminosity at the LDB and age becomes quite shallow. NGC\,2547 is a rich, young
(14$-$55\,Myr) and nearby (400$-$450\,pc) open cluster, that offers excellent
opportunities for exploring the early evolution of angular momentum, magnetic
activity and lithium depletion among low-mass stars (e.g.
\citealt*{jeffries00a}). In this paper we describe an attempt to find the LDB in NGC\,2547. In Sect.\,2 we describe previous observations and age determinations for this cluster. In Sect.\,3 we describe a new $RI_{\rm c}Z$ photometric survey and subsequent spectroscopic observations (using the 2dF fibre spectrograph) to identify low-mass members of the cluster. In Sect.\,4 we analyse the spectra of the cluster candidates and attempt to detect the Li\,{\sc i} 6708\AA\ resonance feature. In Sect.\,5 we describe the location of the LDB and the isochronal and LDB ages of the cluster. Our results are discussed in Sect.\,6.

\section{NGC\,2547: Previous observations and age determinations}

NGC\,2547 ($=$\,C0809-491) is an interesting young cluster and the question of
its age has been scrutinized from two traditional view points: main-sequence turn-off and low-mass star isochrone fitting. \citet{claria82} investigated the high mass population, finding a reddening of $E(B-V)=0.06$, an intrinsic distance modulus of $8.25\pm0.22$ and an age of $57\pm27$\,Myr. \citet{jeffries98} estimate an age of $55\pm25$\,Myr and an intrinsic distance
modulus of $8.1\pm0.1$ based on isochrones and models from \citet*{meynet93} and \citet{schaller92}. These models feature a modest amount of convective overshoot (the same sets of isochrones yield ages of 52\,Myr and 100\,Myr for the $\alpha$\,Per and Pleiades clusters respectively). The major sources of uncertainty here are simply the very small number of stars that define the turn-off, together with their photometric errors and uncertain binary status.

\cite{jeffries98} reported {\em ROSAT} X-ray observations and a
$BVI_{\rm c}$ survey of the cluster, identifying a large population of
X-ray active, low-mass PMS stars. Fits to low-mass
isochrones, derived from the models of \citet{dantona97}, indicated an
age of $14\pm4$\,Myr with a very small spread ($<\pm5$\,Myr). Recently,
\citet{naylor02} have re-analysed the $BVI_{\rm c}$ survey of the
cluster and obtained a better age estimate of 20$-$35\,Myr and an intrinsic
distance modulus of 8.00$-$8.15, using new empirically calibrated, low mass isochrones from \citet{dantona97} and \citet*{siess00}. 

\citet{jeffries02} present results of an investigation of Li depletion
among the K and early M stars of NGC\,2547. The targets in that paper
were selected from the X-ray observations and $BVI_{\rm c}$ photometry
summarised in the previous paragraph and observed with the 2dF spectrograph (a year before the observations described in this paper). They found that for some models of PMS evolution (those of \citealt{baraffe98} and \citealt{siess00}), both the positions in the $V/(V-I_{\rm c})$ colour-magnitude diagrams and the degree of PMS Li depletion for the majority of 0.5-0.9\,M$_{\odot}$ stars could be {\em simultaneously}
satisfied at an age of $\simeq 35$\,Myr. However, these observations
did not look at stars faint enough to put a strong constraint on the
age from the LDB. From the same data, \citet{jeffries00b} reported no
evidence of the Li\,{\sc i} feature in stars as faint as $I_{c}\simeq
16$ and hence estimated an age $>$\,23\,Myr -- quite consistent
with the Li depletion seen in higher mass stars and the nuclear
turn-off age.

\section{Observations and data reduction}

\subsection{$RI_{\rm c}Z$ photometric survey}
\label{photometry}

New photometric data for NGC\,2547 were obtained with the 0.91-m telescope at the Cerro Tololo Interamerican Observatory (CTIO), from 4th to 9th February 1999. The dataset consists of 23 overlapping 13.5\arcmin$\times$13.5\arcmin\ fields in $R$, $I_{\rm c}$ and $Z$. The survey covers an area of approximately a square degree. Individual frames were de-biased, flat-fielded with twilight sky flats and in the case of the $Z$ frames, de-fringed using a median stack of all the night-time $Z$ exposures. The analysis of these images was achieved using the procedures and algorithms described in detail by \citet{naylor02}. Photometric calibration onto the Cousins $RI_{\rm c}$ system was done with reference to 26 measurements of standard stars from \citet{landolt92}. Only two stars with $(R-I_{\rm c})>1.5$\,mag were included and so although the global fit to the standards had an rms discrepancy of 0.04\,mag in $I_{\rm c}$ and 0.03\,mag in $(R-I_{\rm c})$, we believe there may be systematic errors in the colours and magnitudes for $(R-I_{\rm c})>1.5$\,mag of up to 0.1\,mag (for the redder objects we found a systematic effect in I$_{\rm c}$-band magnitude of 0.07\,mag when comparing our catalog with the one presented in \citet{naylor02}). $(I_{\rm c}-Z)$ colours were calibrated using standard stars defined by \citet{zapatero99} in Landolt's SA98 field. The calibrations were taken from the best photometric night. Frames taken on other nights were tied to this calibration via the substantial (2.5\arcmin) overlaps between fields (see \citealt{naylor02}). Our survey reached (S/N$\sim10$) depths of $R\simeq 21.5$, $I_{\rm c}\simeq 20$ and $Z\simeq 20$. The astrometry has been calibrated against the USNO A2 catalogue \citep{monet98}, and yields typical positional accuracies of 0.25\arcsec rms.

\begin{figure*}
\includegraphics[totalheight=11cm]{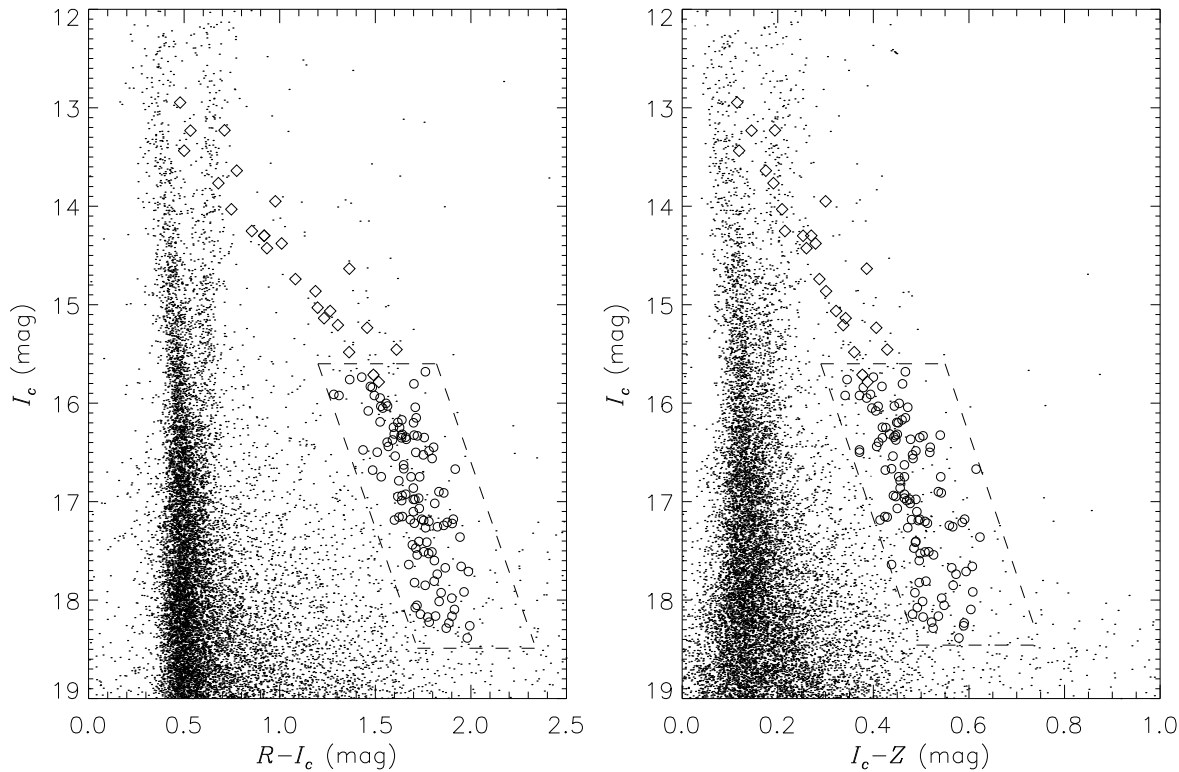}
\caption{$I_{\rm c}/(R-I_{\rm c})$ and $I_{\rm c}/(I_{\rm c}-Z)$
color-magnitude diagrams for the central NGC\,2547 fields. The
pre-main-sequence can clearly be seen separated from the bulk of the
objects. The parallelograms in both diagrams indicate the selection
domain for the follow-up spectroscopy discussed here. The circles
are the objects described in this paper and diamonds are the brighter,
X-ray selected sample discussed in \citet{jeffries02}. }
\label{cmd}
\end{figure*}

Fig.\,\ref{cmd} shows the $I_{\rm c}/(R-I_{\rm c})$ and $I_{\rm
c}/(I_{\rm c}-Z)$ colour-magnitude diagrams of the
$0.6\degr\times0.6\degr$ central area of our survey. The
pre-main-sequence can clearly be seen separated from the back- and
foreground populations. The symbols represent objects with 2dF
spectra: either from the sample discussed here (circles) or from the
earlier observations of X-ray selected brighter members from \citet[][ diamonds]{jeffries02}.

\subsection{2\lowercase{d}F spectroscopic survey}
\label{2df}

The primary goal of these observations was to determine the position of
the LDB by inferring Li abundances from the Li\,{\sc i} 6708\,\AA\
resonance doublet. Observations of cool young PMS stars suggest that
undepleted Li will produce a saturated equivalent width (EW) of $\sim$\,0.6\,\AA\ for this feature \citep{zapatero02}. Theoretical curves of growth tally with these observations and predict that a star depleted by a factor of 100 from its initial Li abundance will still have an  Li\,{\sc i} 6708\,\AA\ EW of $\sim$\,0.3\,\AA\ \citep{zapatero02} but that the line disappears rapidly for even lower abundances (e.g.\ \citealt{pavlenko95}). According to evolutionary model predictions (e.g. \citealt{dantona97,baraffe98}) the Li abundance should change from undepleted to depleted by more than a factor of 100 over a $\sim$\,0.2\,mag increase in luminosity. Thus measurements with an EW accuracy of $\sim$\,0.1$-$0.2\,\AA\ should be capable of identifying the LDB.

Candidate cluster members were selected as spectroscopic targets based
on their positions in {\em both} the $I_{\rm c}/(R-I_{\rm c})$ and
$I_{\rm c}/(I_{\rm c}-Z)$ colour-magnitude diagrams. We were guided by
the predictions of empirically calibrated models and the positions of
previously identified members of both NGC\,2547 and IC\,2391 (a cluster
of comparable age) in these colour-magnitude diagrams (see
Sect.\,\ref{iso_age}). The 109 spectroscopic targets have
magnitudes of 15.6\,$<$\,$I_{\rm c}$\,$<$\,18.3 and are within a
0.8$\degr$\,$\times$\,0.8$\degr$ field. Tables\,\ref{table_li}\,\&\,\ref{table} list the coordinates, I$_{\rm c}$ magnitudes, $(R-I_{\rm c})$ and $(I_{\rm c}-Z)$ colours for the objects observed, as well as a membership flag and  measurements of the lithium feature (Sect.\,\ref{spectral_analysis}). To make it easier for the reader, we have separated the description of the sample into:
Table\,\ref{table_li} for those objects that are suspected to be Li
rich, and Table\,\ref{table} for objects with no detected lithium
(Sect.\,\ref{lithium}). The spectroscopic sample contains about 50 per cent of ``eligible'' targets (those with the right colours and magnitudes) on average, varying from 40 per cent at the brighter magnitudes to 65 per cent for $I_{\rm
c}>17$\,mag. However this must be a lower limit to the percentage of cluster members observed because both the cluster members and the spatial distribution of our targets are centrally concentrated. \citet{littlefair03} use the membership catalogue of \citet{naylor02} to deduce that the half-mass radius for
low-mass stars in NGC 2547 is about 0.2$\degr$. In other words, given
that we expect some contamination by non-members (see Sect.~\ref{contaminate}), we expect more contamination in the outer parts of our survey.
The dense coverage and relatively uniform spread in target magnitudes of our sample should ensure that the LDB, if within this magnitude range, can be located accurately.

Spectra were obtained on the nights of 28th and 29th of February 2000 
with the fiber-feed multi-object spectrograph 2dF \citep{lewis02} at the prime focus of the 3.9\,m Anglo-Australian Telescope (AAT). Each 2dF field plate has 400 object fibers which are 140\,$\mu$m in diameter, corresponding to about
2$\arcsec$ on the sky. This implies that target positions have to be accurate to better than about 0.5$\arcsec$ to avoid significant light loss. To achieve adequate astrometric accuracy, the target positions were cross-correlated with the SuperCOSMOS Sky Survey catalog \citep{hambly01a,hambly01b}. The 2dF
configuration program {\sc configure} was then used to allocate as many
objects as possible to fibers in the 2dF field, subject to constraints on fiber proximity.  About 15 fibers in each field are assigned to sky positions to allow accurate sky subtraction. The final fiber positioning and observational procedures are mostly
automated\footnote{http://www.aao.gov.au/2df/manual/2df\_manual\_inline.html}.
There are two identical spectrographs and CCD detectors, each receiving 200
fibers. Therefore each science exposure actually comprises two individual CCD
images, that are reduced independently. The 1200R grating was used (in both
spectrographs) to give a $\sim$\,2.2\,\AA\ resolution over a range of 1100\,\AA\ centered at 6700\,\AA. 

From the experience of a previous 2dF service run \citep{jeffries00b,jeffries02}, the desired equivalent width accuracy can be achieved with spectra of signal-to-noise $\sim$\,15 per 1.1\,\AA\ pixel. Individual exposures times (per configuration) range from 2700 to 3600\,sec. For most targets multiple exposures were obtained, to achieve a higher signal-to-noise, with total exposure times per target ranging from 1.3\,h to 7.8\,h.

Data reduction was performed using the 2dF Data Reduction system ({\sc 2dfdr} version 2.0 for {\sc linux}), mostly following the advised reduction steps. We did not perform scattered light subtraction or flatfield the spectra, since these procedures were introducing spurious artifacts. Relative fiber transmissions were obtained from offset sky frames and an average sky spectrum was obtained from sky fibers in the stellar frames. Multiple exposures of each field configuration were combined within 2dfdr. Once the reduction  procedure was concluded, the resulting combined frames were analysed using the IDL\footnote{http://www.rsinc.com/idl/index.asp} software.
 
\section{Spectral Analysis}
\label{spectral_analysis}

\begin{figure*}{ht}
\includegraphics[height=9.7cm]{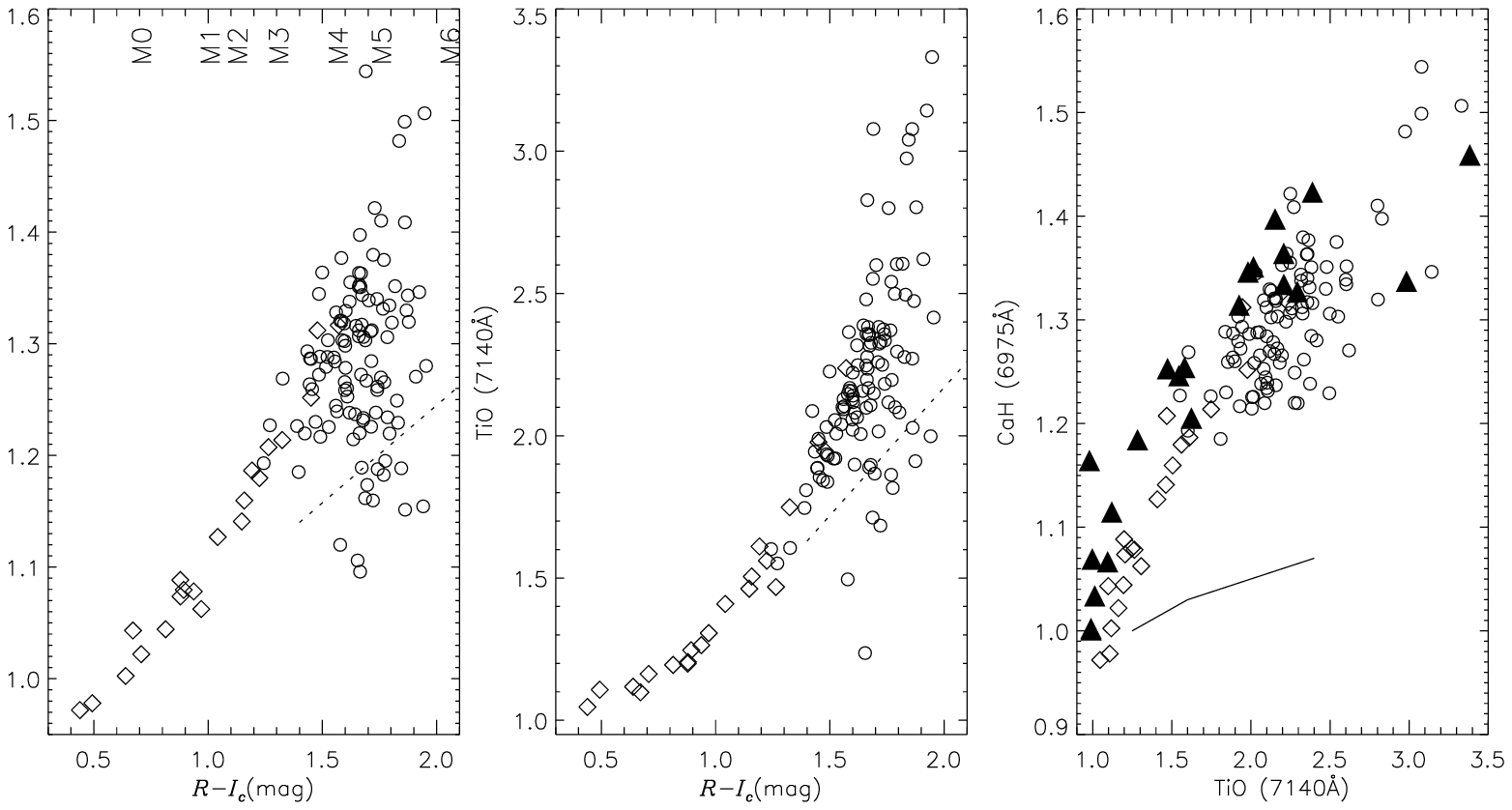}
\caption{(a) CaH\,(6975\,\AA) and (b) TiO\,(7140\,\AA) spectral indices against
$(R-I_{\rm c})$ colour for the sample discussed in the present paper
(circles) and for cluster members from \citet{jeffries02}
(diamonds). The spectral indices check the consistency of the
photometry and observed spectra. We indicate in the left-most plot
approximate colours for disk M-dwarfs \citep{leggett92,leggett96}, assuming
$E(R-I_{\rm c})$=0.04. We retain in our sample only objects that seem to be bona fide late-type dwarfs and cluster members (objects above the two dashed lines). (c) For these objects we plot the CaH against TiO indices together with spectral standards (full triangles,  \citealt{montes97,navascues99}). The solid line is a locus for giants from \citet{allen95}.}
\label{specindices}
\end{figure*}  

\subsection{Optimal variance determination and combination of stellar spectra}
\label{variance}

Most objects were observed several times during the run with different fiber
configurations. In order to better combine these spectra, an accurate
determination of the true variance of each spectra is needed. The major
difficulty is separating the intrinsic noise in the spectrum from the rich
background of spectral features and molecular bands that permeates this spectral region for late spectral types.

Using the difference of pairs of spectra of the same star, we are able to
reasonably remove the ``feature noise'' component and are left only with the
true variance. As expected, the variance of the spectra of the brighter targets follows a linear relation with number of counts and the inverse of the slope measures the effective gain of the CCD camera. The scatter reflects different observing conditions and fiber throughput corrections. Using this linear relation, we can estimate the true variance for each spectrum and thus the true signal-to-noise ratio. Multiple spectra of the same object were averaged, weighted by the calculated variance.

This procedure works less well towards the faintest objects. Sky
subtraction accuracy in the continuum count level is of the order of
6$-$10\% of the original sky level. For the fainter objects, the sky signal
down a fiber is larger than the stellar flux level; so the final stellar flux (in counts) can be rather uncertain. This is taken into account when computing the errors in measured equivalent widths of spectral lines (Sect.\,\ref{lithium}).

\subsection{Spectral indices}
\label{indices}

Our 109 spectra of candidate NGC\,2547 members were analysed firstly by looking at two
spectral indices and comparing those with spectral standards. These spectral
indices measure the strength of the TiO and CaH molecular bands and are
defined as \citep{allen96,briceno98}
\[
\rmn{TiO}(7140 \rmn{\AA}) = \frac{\rmn{C}(7020-7050 \rmn
{\AA})}{\rmn{TiO}(7125-7155 \rmn{\AA})} \, ,
\]
\[
\rmn{CaH}(6975 \rmn{\AA}) = \frac{\rmn{C}(7020-7050 \rmn
{\AA})}{\rmn{CaH}(6960-6990 \rmn{\AA})}\, ,
\]
where C\,(7020$-$7050\,\AA) represents the pseudo-continuum and
TiO\,(7125$-$7155\,\AA) and CaH\,(6960$-$6990\,\AA) molecular absorption bands,
integrated in the indicated wavelength intervals.

\begin{figure*}
\includegraphics[height=11cm]{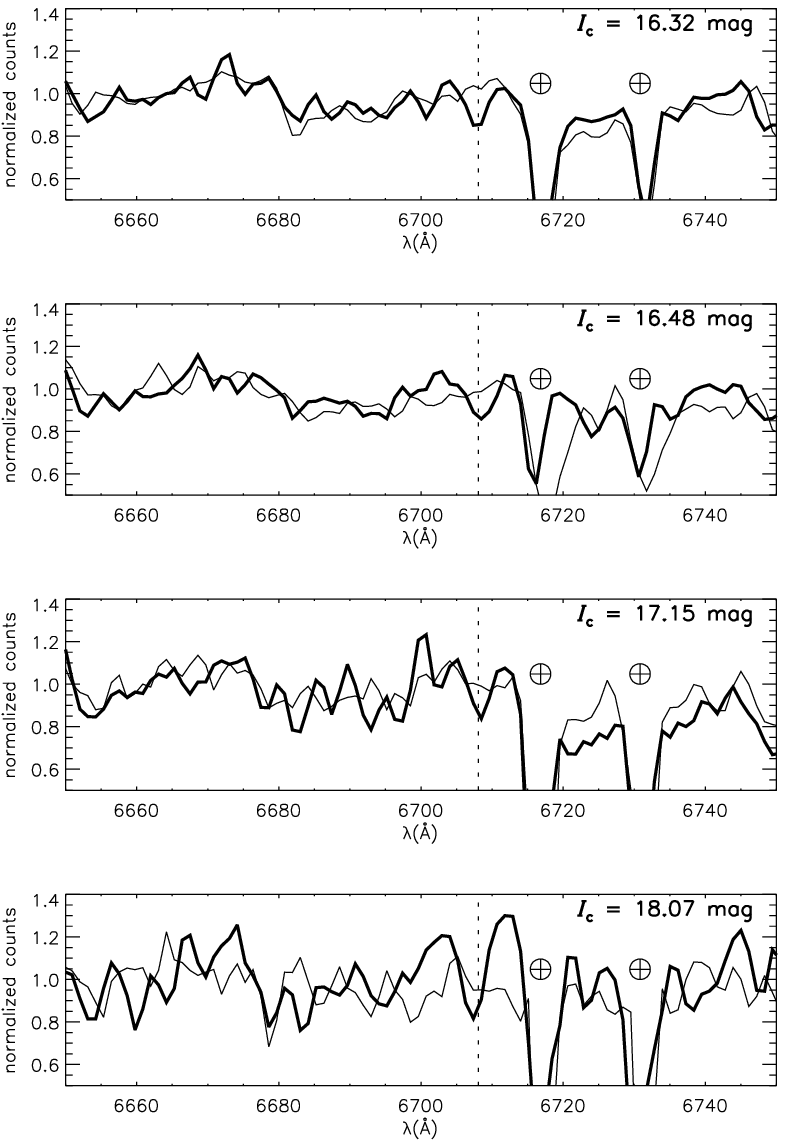}
\includegraphics[height=11cm]{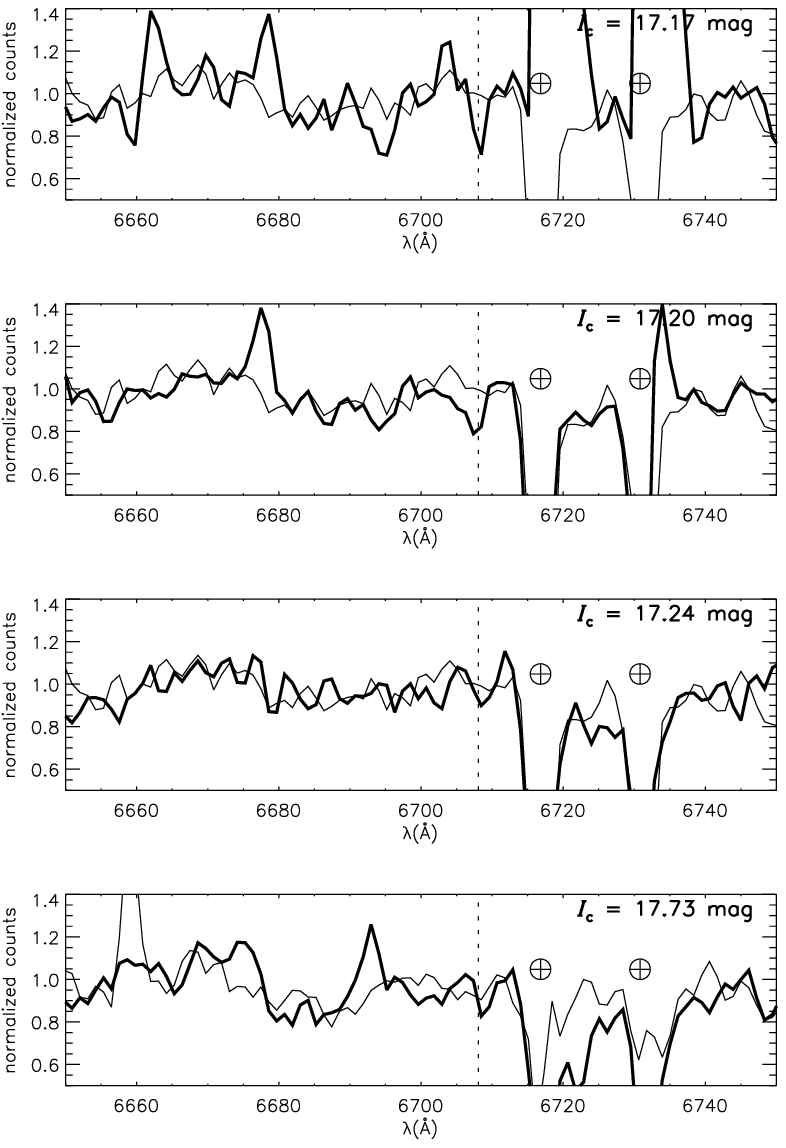}
\caption{Identifications of 6708\,\AA\ Li\,{\sc i} profile in the
sample of NGC\,2547 cluster members. All the profiles were normalised
to the count level in the same wavelength interval in the neighbourhood
of the lithium feature. We have indicated on the top-right corner of each
graph the I$_{\rm c}$-band magnitude of the object. The vertical dashed
line indicates the expected position of the Li feature for cluster
members, with an assumed cluster radial velocity of
$\sim$\,13\,km\,s$^{-1}$ \citep{jeffries00a}. The two strong features
at $\sim$\,6717 and 6730\,\AA\ (indicated by $\oplus$) are likely interstellar [S\,{\sc ii}] lines. The four (thick-line)
spectral profiles on the left are positive lithium detections,
while for the four (thick-line) profiles on the right are more marginal; the light-line profiles are examples of objects that do not show the lithium feature, selected to have I$_{\rm c}$-band magnitude similar to that indicated on the top-right corner of each graph.}
\label{spectra}
\end{figure*} 

In Fig.\,\ref{specindices} (a and b) we plot each of these spectral indices
against $(R-I_{\rm c})$ color for our sample and for the sample from
\citet{jeffries02}. Throughout this paper we assume $E(R-I_{\rm
c})$=0.04, consistent with low reddening determined by
\citet{claria82}. We use these diagrams to further refine the sample by
checking if the targets have spectral indices consistent with what is
expected for M dwarfs; we choose to exclude targets that have the
CaH\,(6975\,\AA) index too low for their $(R-I_{\rm c})$ colour
(indicative of a lower gravity e.g.\ \citealt{allen95}) and some
targets that have a deviant TiO\,(7140\,\AA) index. 
Given the sky subtraction errors in the faintest targets, these
spectral indices are uncertain by 0.05$-$0.1 (for CaH) and 0.1$-$0.2
(for TiO). The adopted selection cuts are therefore quite severe and it
is probable that a handful of genuine members (those with poor sky
subtraction) have been excluded. For the purpose of this paper it is
better to be incomplete than include many contaminating non-members
(see Sect.~5.2).

From the original sample of 109 objects with $I\,\ge\,15.6$\,mag we are left with 94 objects with colours and indices consistent with late spectral types
and cluster membership. Using the TiO\,(7140\,\AA) index we
estimate these objects to have spectral types mostly between M3 and M5
(with a few M2 and M6 objects). The objects rejected as non-members
based on their indices are described with the flag NM (non-member) in
Tables\,\ref{table_li}\,\&\,\ref{table}, otherwise they are flagged M
(members). We plot in Fig.\,\ref{specindices} (c) the CaH\,(6975\,\AA)
index against the TiO\,(7140\,\AA) index for our sample, together with
several standard stars and a locus (from Allen \& Strom
1995), indicating where low gravity giants would lie. Our cluster
candidates have dwarf-like gravities, although there is a hint that the gravities are lower on average than in the field stars. This {\it is} what 
we would expect from a population of M dwarfs
in a cluster of age $\sim30$\,Myr. The models of Chabrier \& Baraffe
(1997) indicate that M-dwarfs of 0.1 to 0.5$M_{\odot}$ and age 30\,Myr, have surface gravities that are between 0.5 and 0.3 dex lower than stars of similar spectral type at $\sim 5$\,Gyr,  but which are still several dex higher than M giant gravities.

In principle, the H$\alpha$ line could also be used as a membership criterion. Young, magnetically active M2$-$M6 PMS stars should exhibit strong chromospheric H$\alpha$ emission, with an EW of 1$-$30\,\AA. Unfortunately, the strength of H$\alpha$ in the sky spectra (arising both from the night sky and an H\,{\sc ii} region apparently centred on NGC\,2547) and the uncertainties in the relative
fiber throughput calibration, mean we are unable to provide a useful
estimate of the H$\alpha$ EW for our targets.

\subsection{The Li\,{\sc i} feature detections}
\label{lithium}

\begin{table*}
\begin{minipage}{138mm}
\caption{Objects with positive or marginal Li identifications. Column\,1 is a list number for purposes of easy cross-identification. Columns\,2$-$9 are objects position and photometry. Column\,10 is a membership flag (NM for non-members and M probable members) and column\,11 the EW with its uncertainty.} 
\label{table_li}
\begin{tabular}{rccllllllcl}
\hline
       & \multicolumn{2}{c}{J2000 Position} & \multicolumn{2}{c}{I$_{\rm c}$-band} & \multicolumn{2}{c}{$(R-I_{\rm c}$)} & \multicolumn{2}{c}{$(I_{\rm c}-Z$)} & & \multicolumn{1}{c}{Li}\\
        & ra & dec & mag & error & mag & error & mag & error & flag & \multicolumn{1}{c}{EW (\AA)}\\
\hline
 26 & 8 12 28.29 & $-$49 16 53.4 & 16.325 & 0.006 & 1.703 & 0.010 & 0.540 & 0.007 & M & 0.486$\pm$0.07\\
 38 & 8 08 16.68 & $-$49 15 35.6 & 16.485 & 0.004 & 1.781 & 0.007 & 0.489 & 0.005 & M & 0.397$\pm$0.09\\
 65 & 8 08 35.99 & $-$49 07 49.7 & 17.151 & 0.004 & 1.642 & 0.006 & 0.424 & 0.005 & M & 0.662$\pm$0.11\\
100 & 8 09 53.16 & $-$49 05 37.8 & 18.076 & 0.017 & 1.709 & 0.030 & 0.492 & 0.019 & M & 0.880$\pm$0.23\\
\hline
 67 & 8 10 13.90 & $-$49 39 48.2 & 17.178 & 0.008 & 1.908 & 0.014 & 0.591 & 0.009 & M & 0.555$\pm$0.27\\
 72 & 8 08 29.18 & $-$49 03 54.9 & 17.201 & 0.016 & 1.780 & 0.027 & 0.508 & 0.016 & M & 0.404$\pm$0.14\\
 75 & 8 09 50.22 & $-$49 21 15.8 & 17.241 & 0.016 & 1.858 & 0.025 & 0.559 & 0.016 & M & 0.334$\pm$0.21\\
 91 & 8 10 46.12 & $-$49 06 28.5 & 17.737 & 0.017 & 1.824 & 0.059 & 0.573 & 0.020 & M & 0.253$\pm$0.15\\
\hline
\end{tabular}
\end{minipage}
\end{table*}

Our main goal is to determine the position of the lithium depletion
boundary for NGC\,2547. To that end we have analysed the spectra of the
94 remaining cluster candidates covering the range 15.6$-$18.3\,mag in $I_{\rm c}$.

In Fig.\,\ref{spectra} we plot the Li\,{\sc i} profiles we found in our
target sample. The plots on the left are (we claim) positive detections of the lithium feature, while on the right are more marginal detections (all shown as thick-lines); we have overplotted (with light-lines) example spectra of objects with similar $I_{\rm c}$ magnitudes that appear not to show the Li feature.

For the 8 objects in Fig.\,\ref{spectra} we give their Li
equivalent width\footnotemark[3] (EW) and errors
(Table\,\ref{table_li}). The Li feature was integrated over a fixed 5
pixel interval, corresponding to a width of 5.5\AA. The EW error has
two components that are added in quadrature. The error due to pixel-to-pixel
uncertainties on the flux values within the integration interval can be described as $\delta$EW\,=\,$(\sqrt{rp})$/SNR, where r and p are the width of the feature and the pixel width respectively, and SNR is the signal-to-noise
ratio per pixel (as derived in Sect.\,\ref{variance}). The other error
contribution comes from the uncertainties in the continuum flux level
(Sect.\,\ref{variance}) from the estimated sky subtraction error.
The latter error is usually negligible, except in the
faintest objects. When no feature is detected, the detection threshold is given by  a total  2-$\sigma$ upper limit in Table\,\ref{table}. 
\footnotetext[3]{In reality we are measuring a pseudo equivalent width 
with respect to the local pseudo-continuum (e.g.\, \citealt{zapatero02})}

\section{How old is NGC\,2547 ?}

An LDB location within the magnitude range we have analysed corresponds
to an age range of $\sim$\,20$-$50\,Myr. We have detected lithium in a
few cluster members but have we detected the LDB? The lithium abundance
drops from undepleted to 99 per cent depleted in just over 0.2\,mag, therefore
we expect to find objects with either entirely depleted or undepleted
Li abundances, with no (or very few) transition objects. Initially it
would seem that we have not found a clear boundary in our dataset and 
so have not found the LDB in NGC\,2547. There are several possibilities
to explain this: {\it i}) the data were not of sufficient quality to detect the LDB; {\it ii}) we did not observe any cluster members; {\it iii}) the cluster
is older than 50\,Myr and our sample is not deep enough to detect the
LDB.  In this section we discuss the limits of our spectroscopic
sample, try to constrain the cluster (LDB and low-mass isochronal) age and
compare with IC\,2391, a cluster previously thought to be $\sim5$\,Myr
older than NGC\,2547 \citep{jeffries98}.

\subsection{How faint can we detect the Li feature?}

\begin{figure}
\includegraphics[height=6.cm]{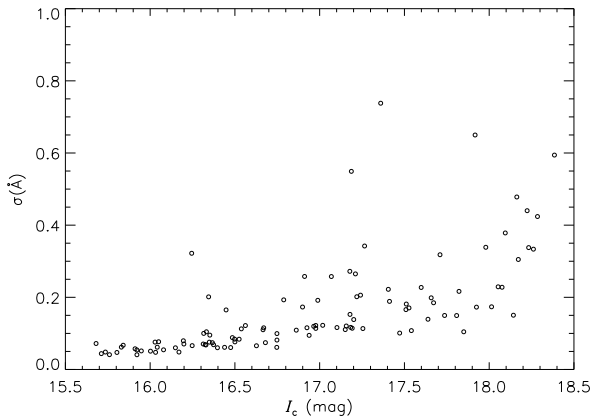}
\caption{Total EW errors versus $I_{\rm c}$ for the cluster candidates in the sample. }
\label{errorplot}
\end{figure}

The total EW errors are plotted versus $I_{\rm c}$ in
Fig.\,\ref{errorplot}. The scatter in the relationship is largely due
to variations in fiber transmission efficiency.
If we expect Li-undepleted stars to have EWs of
around 0.6\,\AA\ (e.g.\, \citealt{zapatero02}), then we judge that we
are capable of detecting (at 3-sigma confidence) individual Li-rich
objects to $I_{\rm c}\simeq 17.5$\,mag for the majority of targets and a
{\em population} of Li-rich objects (at a lower individual significance
of say 2-sigma) to $I_{\rm c}\simeq 17.8$\,mag.  Thus, providing our sample
is not very heavily contaminated by cluster non-members down to these
magnitudes we should have been capable of detecting the LDB.

\subsection{Have we observed cluster members?}

\label{contaminate}

A first point to make is that it is extremely unlikely that we have
``missed'' the NGC 2547 PMS. A glance at Fig.~\ref{clean_cmd_compare}
shows that even though we assumed an age of $\sim$20$-$50\,Myr in making
our candidate selection on the CMDs, the isochrones are so closely
packed together compared with the width of the selection area that no
plausible error in our assumed age would allow the PMS to fall outside
this area.

Another important consideration is at what level is our sample contaminated by
cluster non-members? By
using two colour-magnitude diagrams to select pre-main-sequence objects and by
further refining that sample by analysing their spectral indices, we have
removed background giants and other anomalously reddened
objects. However our final sample can still contain field M dwarfs that
are basically indistinguishable from cluster members on the basis of
their colours and spectral indices. As we expect almost all of these to
have completely depleted their lithium, it is relevant to estimate 
the contaminant numbers. 

As the field-of-view towards the cluster is largely unobscured we ignore the
effects of reddening. We adopt the model of \citet{ortiz93} that describes star
counts at red and infrared wavelengths in the solar neighbourhood. For
late spectral types, the dominant component is the so-called ``thick
disk'' component characterised by a scale height of 390\,pc (i.e.\ the
old disk population). Taking into account the galactic latitude of
NGC\,2547 ($|b|\,\sim$\,9$\degr$) we can estimate the stellar density
for each spectral type. \citet{leggett92} and \citet{leggett96} list
typical $(R-I_{\rm c})$ and M$_{\rm I_{\rm c}}$ for each spectral
type. From the $I_{\rm c}/(R-I_{\rm c})$ colour-magnitude diagram we
estimate the distance range over which a M-dwarf with a given
$(R-I_{\rm c})$ (spectral type) would appear to populate
the parallelogram in Fig.~\ref{cmd}. Then, given the {\it effective} angular extent for our spectroscopic survey (recall that we only observed between 40
and 65 per cent of ``eligible'' targets -- Sect.~3.2)  and assuming the spectroscopic targets are spread evenly over this area (see below), this defines a contaminating volume that together with the stellar density allows us to compute the number of contaminants. 

We estimate the level of contamination to be 20, 30 and 10 per cent
respectively for spectral types M3, M4 and M5 (the largest contaminating volume occurs for spectral type M4). Our targets show some concentration towards the cluster centre, as expected given the small 0.2$\degr$ half-mass cluster radius. In addition, Fig.~\ref{cmd} shows
that our targets are concentrated towards the centres of the
parallelogram selection domains, whereas foreground contaminants would
tend to be concentrated towards the fainter edge at a given colour.
Given this, it is likely that these contamination numbers (calculated
assuming a uniform target distribution) are overestimates. However, to
be absolutely conservative, we assume a uniform contamination of 30 per cent
across the sample spectral type range. Thus the majority of the
spectroscopically selected cluster candidates should indeed be cluster
members and contamination by field M dwarfs should not greatly impair
our ability to detect the LDB.

\begin{figure}
\includegraphics[totalheight=12cm]{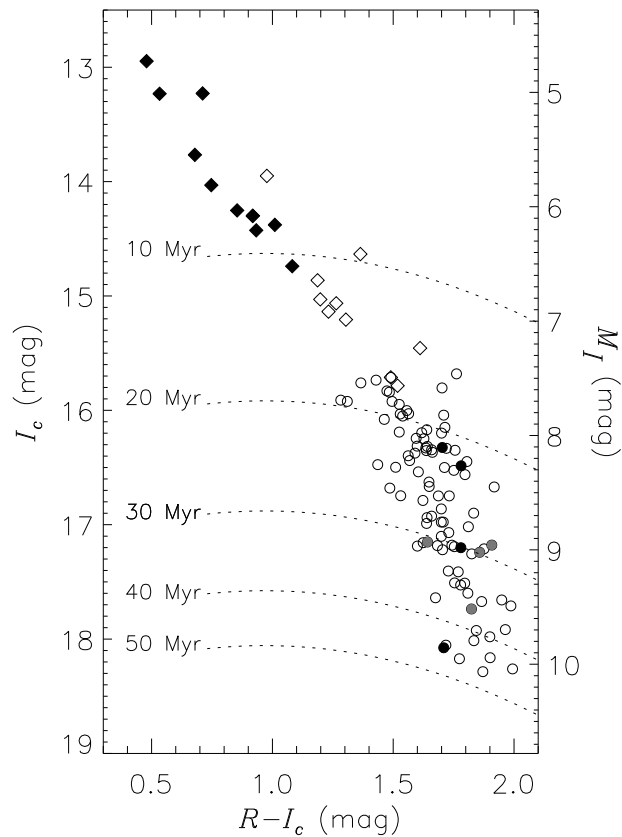}
\caption{$I_{\rm c}/(R-I_{\rm c})$ colour-magnitude diagram of the
spectroscopically observed cluster candidates. Symbols are as in Fig.\,\ref{cmd}; filled symbols are objects with detected lithium and grey symbols are objects with marginal detections (see Sect.\,\ref{lithium}). We adopt a distance modulus and reddening respectively $(m-M)_{0}$\,=\,8.15 and
$E(R-I)$\,=\,0.04. The dashed lines are isochrones of 99 per cent lithium depletion using the models of \citet{chabrier97}.}
\label{clean_cmd}
\end{figure} 

\subsection{The location of the LDB}
\label{ldb}

Fig.\,\ref{clean_cmd} shows the $I_{\rm c}/(R-I_{\rm c})$
colour-magnitude diagram of the cluster candidates. The filled symbols
represent lithium-rich objects and grey symbols are the more
marginal detections. The dashed lines are isochrones of 99 per cent lithium
depletion, based on the \citet{chabrier97} models and described in the
next section. We have adopted a conservative distance
modulus of $(m-M)_{0}$\,=\,8.15 and reddening $E(R-I)$\,=\,0.04
\citep{naylor02}. We divide the discussion of lithium detection and the
location of the
LDB into $I_{\rm c}$-magnitude bands: $I_{\rm c}$\,$<$\,17.2\,mag;
17.2\,$\la$\,$I_{\rm c}$\,$<$\,17.8\,mag and 17.8\,$\la I_{\rm
c}$\,$<$\,18.3\,mag.

There are two objects at $I_{\rm c}\simeq16.5$ with detected lithium
(Fig.\,\ref{spectra}). We have plenty of good quality spectra for
fainter objects and with only an estimated 30 per cent
contamination, we can safely say that the LDB is not
at $I_{\rm c}<$\,17.2\,mag. What are these lithium-rich objects? A few such objects, brighter than the LDB and lithium-rich, have also been found in comparable samples in the Pleiades \citep{oppenheimer97} and IC\,2391 \citep{navascues03}. It has been suggested that such objects might belong to a later burst of star formation or might not have formed in these clusters, but no convincing explanation is available. Both objects in NGC 2547 could be binary systems, based on their position in the colour-magnitude diagram, but could also be isolated very young ($<10$\,Myr) stars unassociated with NGC\,2547.

At $I_{\rm c}\simeq 17.2$\,mag there is a lithium-rich object and a few tentative detections. Again, based on the fact that there are good quality
spectra to magnitudes as faint as 17.8\,mag and that none of those
objects shows a convincing lithium detection, we are inclined to believe
that we are not seeing the LDB at $I_{\rm c}\simeq$\,17.2\,mag. Taking into account that if present the lithium feature should be strong (EW$\sim0.6$\,\AA), then the tentative detections are not at all convincing. The definite lithium-rich object could be a binary system, in which case the true LDB could be up to 0.75\,mag fainter, although its position in the colour-magnitude diagram suggests 0.3-0.5 mag fainter might be more plausible. Taking into account the sensitivity of our spectra we thus believe the LDB for NGC\,2547 is not at $I_{\rm c} <$17.8\,mag.

Most of the spectra of objects with $I_{\rm c}\ga$17.8\,mag are of rather
poor quality, with the exception of the Li-rich object with $I_{\rm c}\,=\,18.07$\,mag in Fig.\,\ref{spectra}. Thus individually these noisy spectra do not allow us to locate the LDB.  But, given our relatively low estimate of the contamination by non-members, combining these low signal-to-noise spectra seems a reasonable step to take.

\begin{figure}
\includegraphics[totalheight=12.2cm]{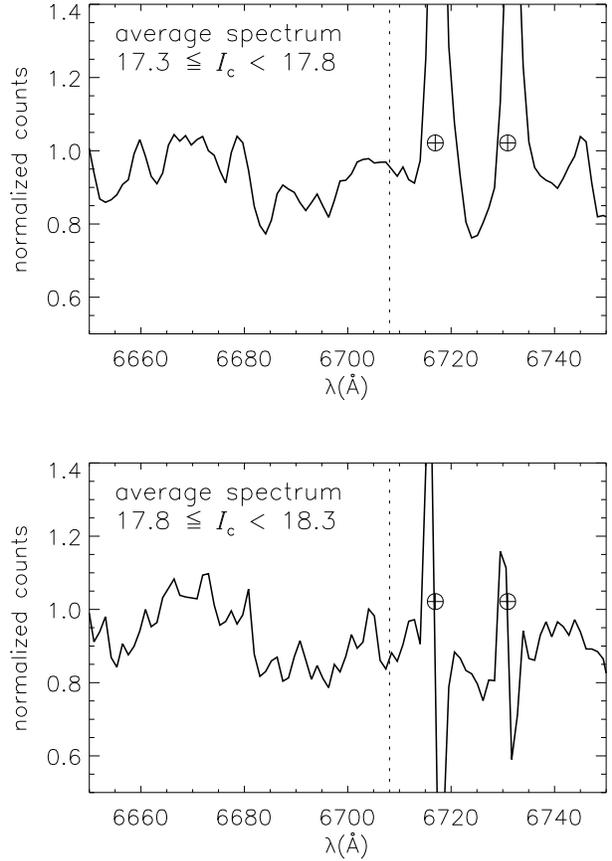}
\caption{Average lithium spectra: on top for objects with
17.3\,$\leq$\,$I_{\rm c}$\,$<$\,17.8\,mag and at the bottom with
$I_{\rm c}$\,$\geq$\,17.8\,mag. For these weighted averages we have
excluded spectra with lithium detections. The first (top) magnitude bin
contains 10 objects ($\sim$\,3 possible contaminants) and the second
(bottom) 9 objects (also $\sim$\,3 possible contaminants).}
\label{average_spectra}
\end{figure} 

In Fig.\,\ref{average_spectra} we have plotted the weighted average spectra of
the objects in the magnitude bins 17.3\,$\le I_{\rm
c}$\,$<$\,17.8\,mag (top graph) and $I_{\rm c}$\,$\ge$\,17.8\,mag (bottom graph). We have excluded the two objects with probable or possible lithium detections from these averages. In the brighter average spectrum the lithium feature is definitely not present (EW$<0.12$\,\AA\ at 2-sigma confidence). 
The average spectrum of the fainter objects (bottom graph) {\it does} show a lithium feature with EW$=0.4\pm0.1$\,\AA\ -- precisely what we would expect if the fainter sample consists of two thirds of stars with an undepleted Li EW
of about 0.6\,\AA\ and one third of stars (contaminants) with no Li feature at all. This points to the LDB being located somewhere between $17.8\le I_{\rm c} < 18.3$.

As the SNR of the spectra is not high, we made a separate test of our uncertainty estimates and ability to detect Li features; we examined all 94 candidate member spectra in a wavelength region (6740$-$6780\,\AA) where no atomic lines are identifiable and where the molecular bands have similar shape
and strength as in the vicinity of the Li {\sc i} 6708\,\AA\ feature. On average we found 1 detected ``line'' per 4\,\AA\ interval (the interval over which we would allow a possible feature to be identified with Li at 6708\,\AA) at the
the 2 to 3-$\sigma$ level with EWs of 0.3$-$0.4\,\AA, which is just what
one would expect from 94 trials. Comparing these statistics with Table\,\ref{table_li} it seems likely that all of the positive detections are real, but one or two of the more marginal detections could be spurious. The numbers of spurious features detected in our test, together with their measured EWs and uncertainties, lends support to our estimates of the significance and strength of the Li features in our targets.

\subsection{The age of NGC\,2547}

\subsubsection{The LDB age of NGC\,2547}

\begin{figure*}
\includegraphics[height=12cm]{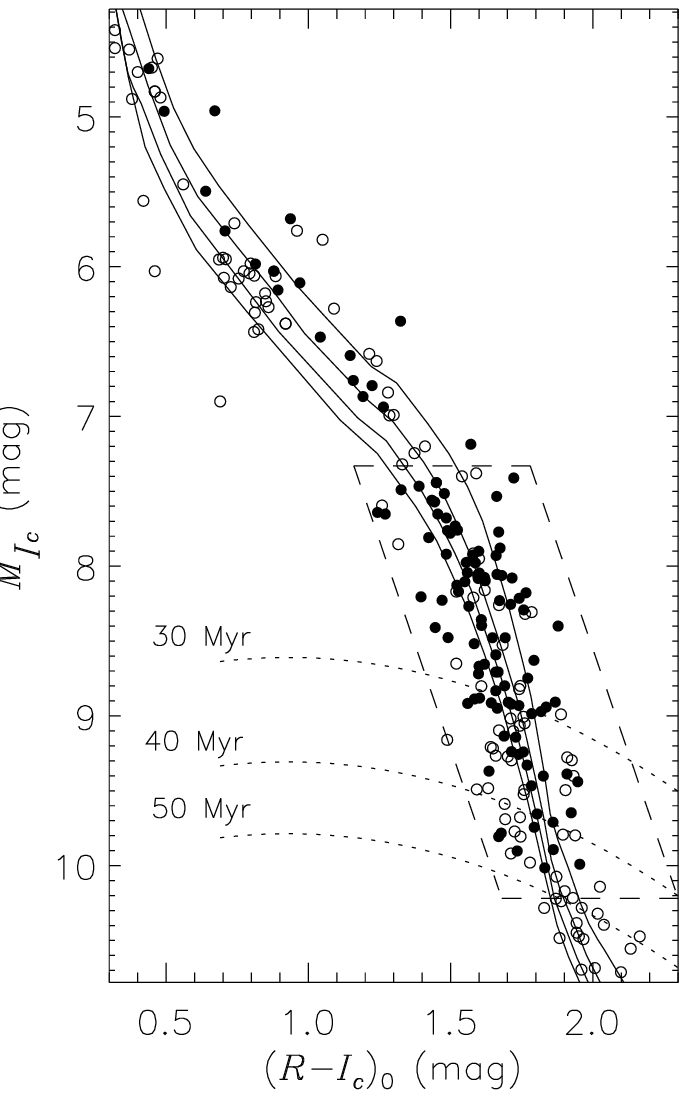}
\includegraphics[height=12cm]{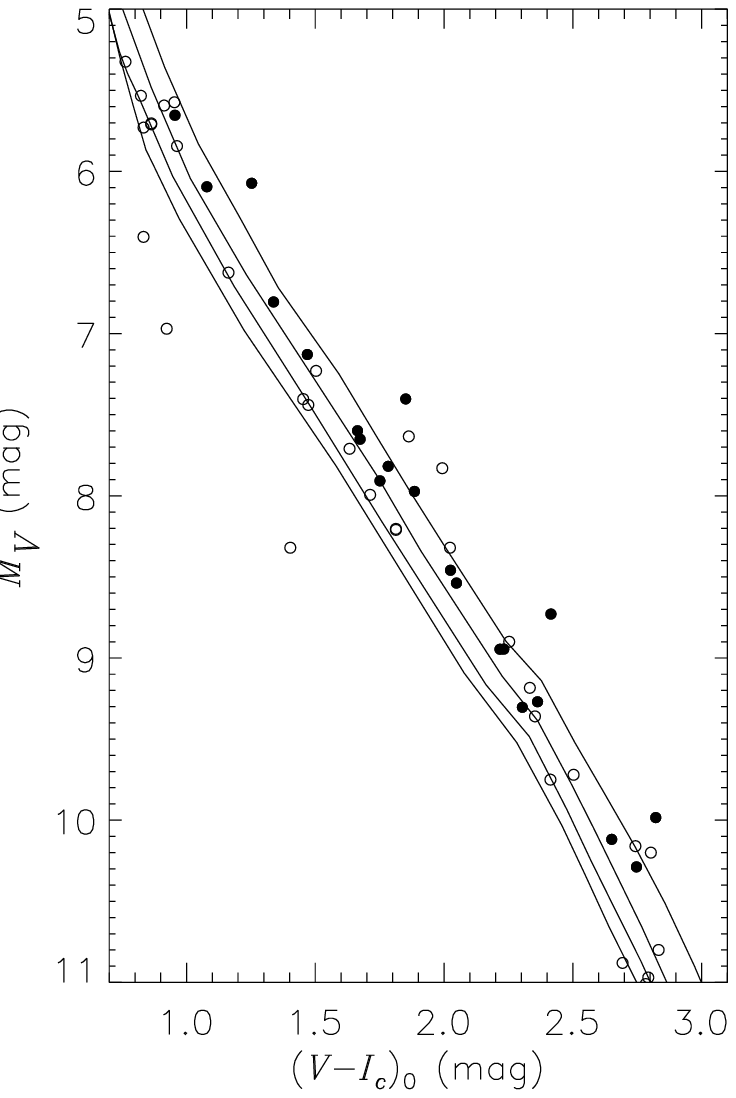}
\caption{$M_{I_{\rm c}}/(R-I_{\rm c})_{0}$ and $M_{V}/(V-I)_{0}$ diagrams for
NGC\,2547 and IC\,2391 cluster members. Filled circles are the
NGC\,2547 members. The IC\,2391 members (open circles) are from
\citet{navascues01}, \citet{simon98} and \citet{patten96}, for a
distance modulus of $(m-M)_{0}$\,=\,5.95 and $E(B-V)$\,=\,0.06. The box
in the diagram is the selection space of the spectroscopic sample. We
only have $VI_{\rm c}$ photometry from the sample of brighter cluster
members, from \citet{jeffries02}. Also represented are LDB isochrones
(see caption Fig.\,\ref{clean_cmd}) and empirically calibrated 20, 30, 40
and 50\,Myr isochrones (full lines), based on the Chabrier \& Baraffe
(1997) models.}
\label{clean_cmd_compare}
\end{figure*}

We assume here that the LDB corresponds to the brightest star whose lithium
abundance has been 99 per cent depleted from its original value. The
(absolute) I$_{\rm c}$-band magnitude of the LDB can be converted to a
cluster age with the help of evolutionary models. One can use
theoretical evolutionary models that incorporate model atmospheres as appropriate boundary conditions and also predict the magnitudes and colours for low-mass objects. These directly yield the I$_{\rm c}$-band magnitude of the LDB at a given age (e.g. \citealt{baraffe98}; \citealt{stauffer98}). However, this is the only set of models that produces such self-consistent magnitudes and colours. Alternatively, \citet{jeffries01b} and \citet{stauffer99} use the observed $I_{\rm c}$-magnitude and $(R-I_{\rm c})$ colour of the LDB together with an empirical bolometric correction-colour relation from \citet{leggett96}
to compute the bolometric luminosity of the LDB. This approach does not use effective temperature-colour relationships or bolometric corrections provided by models. Using any evolutionary model, the LDB luminosity is then converted to an age. This method allows us to perform the same analysis using several different evolutionary models and quantitatively compare the results (see Table\,\ref{ldbtable}). However, for the record, we find that for NGC\,2547, the ages determined by either method are perfectly consistent. The LDB isochrones (dotted lines) in Figs.\,\ref{clean_cmd} and \ref{clean_cmd_compare} were computed following the latter procedure, using the evolutionary models of \citet{chabrier97}.

We have calculated the LDB age assuming that the boundary lies between
$17.8\leq I_{\rm c} < 18.3$ with a corresponding colour of $1.8\leq
R-I_{\rm c} \leq 1.9$. We adopt $E(R-I_{\rm c})=0.04$ and an intrinsic distance
modulus of 8.15. The luminosity at the LDB was then used to estimate
the cluster age using the models of \citet{chabrier97},
\citet{dantona97} and \citet[][ using both Z\,=0.02 and Z\,=0.01 models]{siess00}. The results are presented in
Table~\ref{ldbtable} and are very similar for the differing models.
To estimate the uncertainties we assume that the
location of the LDB is in error (due to {\it systematic} uncertainties
in the photometric calibration) by $\pm0.1$ in colour and magnitude.
The effects of uncertainties in the reddening and bolometric
correction-colour relation will be minor in comparison (see \citealt{jeffries01b}). There is a further uncertainty associated with the
adopted distance. We have used the largest probable distance modulus
found by \citet{naylor02}. In fact the distance modulus could be as low as
8.0, or perhaps even a little lower if NGC 2547 has a significantly
sub-solar metallicity. Using a distance modulus of 8.0 would {\em add}
about 3\,Myr to all the ages in Table~\ref{ldbtable}. Thus the LDB age for
NGC 2547 is at least 35\,Myr (D'Antona \& Mazzitelli models) but
probably less than 54\,Myr (Siess et al. $Z=0.01$ model, distance
modulus of 8.0).

\begin{table*}
\caption{ The LDB and isochronal age for NGC 2547 calculated using different models.}
\begin{tabular}{cccccc}
\hline
 &  & Chabrier \& Baraffe & D'Antona \&
Mazzitelli & Siess et al. $Z=0.02$ & Siess et al. $Z=0.01$\\
\hline
$I_{\rm c}$ & $(R-I_{\rm c})$ & \multicolumn{4}{c}{LDB age (Myr)} \\
17.8 & 1.8 & $38.5\pm2.1$ & $37.1\pm2.1$ & $39.1\pm2.0$ & $39.0\pm2.3$
\\
18.3 & 1.9 & $46.4\pm3.0$ & $44.5\pm2.7$ & $47.1\pm2.8$ & $48.1\pm3.1$
\\
\hline
&&\multicolumn{4}{c}{Isochronal age (Myr)} \\
&&$25\pm5$ & $25\pm5$ & $30\pm5$ & $30\pm5$\\
\hline
\end{tabular}
\label{ldbtable}
\end{table*}

\subsubsection{The isochronal age of NGC\,2547 and comparison with IC\,2391}
\label{iso_age}

In the $M_{I_{\rm c}}/(R-I_{\rm c})_{0}$ and $M_{V}/(V-I)_{0}$ colour-magnitude
diagrams in Fig.\,\ref{clean_cmd_compare}, we compare the NGC\,2547
sample (filled circles) with members of IC\,2391 (empty circles, from
\citealt{navascues01}; \citealt{simon98}; \citealt{patten96}), thought
to be $\simeq5$\,Myr older than NGC\,2547 (we assume
$(m-M)_{0}$\,=\,5.95 and $E(B-V)$\,=\,0.06 for IC\,2391). 

We try to determine the ages of these clusters by using
low-mass PMS isochrones. For NGC 2547 \citet{naylor02} used the evolutionary 
models of \citet{dantona97} and \citet[][ -- the $Z=0.02$ models]{siess00}.  They found an age of 25\,$\pm$\,5\,Myr and intrinsic distance modulus of 8.00$-$8.15 from the D'Antona \& Mazzitelli models and corresponding values of
30\,$\pm$\,5\,Myr for the Siess et al. models. We have repeated the analysis
in Naylor et al. using the models of \citet{baraffe98, baraffe02} which
feature a convective mixing length of 1.9 times the pressure scale
height. The optical colours predicted by these models are known to be
inaccurate (see \citealt{baraffe98}), so we adopt a empirical
calibration procedure (described in detail by \citealt{jeffries01a, naylor02})
to convert from bolometric luminosities and effective temperatures to
magnitudes and colours. Briefly, we assume that the Pleiades has an
intrinsic distance modulus of 5.6 and age of 120\,Myr \citep{stauffer98}. Pleiades photometry is then used along with a bolometric
correction-colour relation to {\em define} a colour-effective
temperature relation that can then be used to produce an isochrone at
any age where the same colour-effective temperature relation is assumed
to apply. 
\nocite{baraffe02}

The solid lines in Fig.\,\ref{clean_cmd_compare} show the results of
this procedure. As we believe that photometric calibrations might be
significantly affected by systematic errors (Section\,\ref{photometry})
for stars with $(R-I_{\rm c}) > 1.5$\,mag (i.e. most of the faint sample),
we attach most weight to the brighter stars ($M_{I_{\rm
c}}\,\la\,7.5$\,mag) and the $VI_{\rm c}$ magnitudes to estimate the
absolute cluster age. For an intrinsic distance modulus of 8.15, an age
of $25\pm5$\,Myr is indicated for NGC\,2547. IC\,2391 appears to be a
little older in both diagrams, perhaps by 5$-$10\,Myr, but after taking into
account the likely distance uncertainties this difference is not very
significant.

The reader might question the empirical calibration based upon Pleiades
photometry, because we are forced to assume a distance and age for the
Pleiades. In fact, our results are quite robust to these
assumptions. We have generated model isochrones using the more recent
(but disputed -- see \citealt{pins98}) Hipparcos-measured
distance modulus of 5.36 \citep{robichon00};  using a Pleiades age of 150\,Myr -- which is the LDB age that would be derived if the Hipparcos distance is adopted \citep{jeffries01b}; and simultaneously adopting both of the
latter changes. We find that changing the age of the Pleiades by
30\,Myr has {\em no} significant effect on our derived isochronal ages simply
because 120\,Myr old stars are already almost on the ZAMS for stars at
the relevant colours. Changing the distance of the
Pleiades simply changes the distance we find for NGC 2547 by a similar
amount. Thus, if we accept the Hipparcos parallax distance to the
Pleiades, then the distance modulus of NGC\,2547 is $7.75-7.90$, the
isochronal ages are unchanged, but the LDB ages reported in
Table~\ref{ldbtable} would be {\it older} by a further $\sim5$\,Myr.

Finally we have independently checked our results by adopting an
empirical relationship between $(V-I_{\rm c})$ and effective temperature
found by fitting data from \citet{bessell79} and \citet{leggett96}. We find
that the low-mass isochronal age for NGC\,2547 would be reduced by about
5\,Myr for all the models we have considered, but the distances are
unchanged.

\section{Discussion}

The main finding of this paper is that while the LDB age of NGC\,2547
lies between 35$-$54\,Myr and is formally consistent with the (rather
uncertain) nuclear turn-off age of $55\pm25$\,Myr, the age found from fitting low-mass isochrones, using the same evolutionary models is lower -- in the range
20$-$35\,Myr. This discrepancy is even more significant than it might
first appear because: (i) the evolutionary models that give the youngest
LDB age also give the youngest isochronal age; (ii) the conservative distance modulus assumed to obtain the LDB ages is larger than the distances required to
fit the isochrones and thus the LDB ages should probably be increased
by a few Myr.

The same appears to be true for IC\,2391. \citet{navascues01}
find an LDB age of $53\pm5$\,Myr. Using the models and techniques
presented in this paper (but Barrado y Navascu\'{e}s et al.'s estimate
of the magnitude and colour of the LDB), we would estimate an LDB age of
48$-$50\,Myr, with a $\pm5$\,Myr uncertainty due to the LDB placement
and systematic errors in the photometric calibrations. The low-mass
isochronal age is not quite so easy to determine as for NGC 2547,
possibly because the available photometry is less precise, but we
estimate 25$-$40\,Myr using the same distance and range of evolutionary models.

Could the discrepancy between isochronal and LDB ages, assuming a
universal $T_{\rm eff}$-colour relationship, mean that there are
systematic problems in the current generation of low-mass evolution
models? Or, could the $T_{\rm eff}$-colour relationship change
sufficiently with gravity between 30 and 120\,Myr to increase the isochronal
ages of the low-mass stars?

We think the latter possibility unlikely, simply because the gravities
of the stars which define the low-mass isochronal ages (stars with
masses of approximately 0.3-1.5\,$M_{\odot}$) do not change
significantly between 30 and 120\, Myr.  The atmospheric models of
\citet{baraffe98} suggest that any age dependence of the $T_{\rm
eff}$-colour relationship is indeed very small.  However, the optical
colours, particularly $(V-I)$ and $(R-I)$, in these models are still not
capable of reproducing the observed colour-magnitude diagrams of young
clusters, probably due to missing sources of opacity in the optical
region.  There is presently insufficient published near IR data in NGC
2547 to do a similar test in a spectral region where the model colours
are likely to be more realistic. However, the available $(I_{\rm
c}-K_{s})$ data for\,IC 2391 yields an identical LDB age but
spectroscopically confirmed members of IC 2391 also clearly lie
{\em above} a 50\,Myr isochrone (and therefore at younger ages) in the
$K_{s}/(I_{\rm c}-K_{s})$ diagram (see \citealt{navascues03}).

We are then left contemplating what physical ingredients may need
altering in the PMS models which might bring the
LDB and isochronal ages into agreement. The different evolutionary
models considered in this paper already cover a wide variety of
treatments of convection, equations of state and atmospheric structure
in low-mass stars. None of these seems capable on their own of closing 
the discrepancy we have identified; in particular, the LDB age is
very robust to changes of detailed model physics. Perhaps then the answer is
that something is missing from the models entirely. Two possibilities
are rotation and dynamo-generated magnetic fields, both of which are expected to be present in young stars. 

Investigation of these factors on the LDB ages are still at a
preliminary stage. Intuitively we would expect that rotation and
magnetic fields provide additional pressure terms, decreasing the core
temperature and delaying the onset of Li burning. This is borne out by
preliminary calculations. Rapid rotation may {\it increase} the LDB
ages by 20$-$25 per cent at 120\,Myr, but by only 1$-$3\,Myr at 30\,Myr
\citep{burke00}. D'Antona (private communication) finds that the
lowering of the core temperature thanks to the extra support provided
by a plausible magnetic field also increases the LDB age. These effects
would seem to be the opposite of what we require, but it is possible
that these physical changes could simultaneously provide a {\em
bigger} increase in the derived isochronal ages because the effective
temperatures are also altered.

\citet{pins98} have shown that rapid rotators lie up to 0.1
mag above a $V/(V-I_{\rm c})$ isochrone defined by the slower rotators in the
Pleiades and $\alpha$\,Per open clusters. If almost all
the low-mass stars in NGC\,2547 are fast rotators this might mean the
low-mass isochronal ages were underestimated, but only by $\simeq$5\,Myr.
The magnetic field hypothesis may hold more promise. The effect in the
Hertzsprung-Russell diagram is to move a star of a given mass and age
to significantly cooler temperatures at roughly constant luminosity -- making them appear much younger when interpreted with models featuring no magnetic
field \citep{dantona00}. 

If either rotation or magnetic fields can reconcile the LDB and
isochronal ages, the implication is that the cluster age scale becomes
even older than suggested from the present group of LDB ages, i.e.\,
even more discrepant with nuclear turn off ages from models featuring
no convective core overshoot.

Further clues may arise from PMS associations where there are
independent kinematic age indicators.  \citet{song02} have found that while
the lower-mass component of a recently discovered PMS binary system (GJ
871.1B) exhibited Li, the higher mass component (GJ 871.1A) did
not. They deduced that if the lower mass component was at or beyond the
LDB, the system would have an age $>20$\,Myr (we find $25\pm3$\,Myr
based on the procedures and range of models described in this
paper). However, using assumed colour-effective temperature relations,
bolometric corrections (derived for old disk M dwarfs) and a Hipparcos
parallax, they find the isochronal age of the system is $\le$\,10\,Myr.
Furthermore Song et al. claim that this binary may be a member of the
$\beta$ Pictoris moving group which has both a low-mass isochronal age and
kinematic expansion age of $\simeq12$\,Myr. If this were the case (and
we note that expansion ages measure the time since an association
became unbound and therefore are lower limits to the true age),
then any scope for increasing the isochronal or LDB ages is severely
limited and a way must be found to {\em increase} the rate at which Li
is burned in very low mass stars.

\section{Summary}

We have obtained intermediate resolution 
2dF spectra of many candidate low-mass members of
NGC\,2547. We have refined our sample to select 94 objects with
$(R-I_{\rm c})$ and $(I_{\rm c}-Z$) colours  
and narrowband spectroscopic indices that are
consistent with cluster membership.
Our goal was use the Li\,{\sc i} feature as an age indicator,
by detecting the lithium depletion boundary of this cluster. We were
not able to conclusively find the position of such a boundary in the
$I/(R-I_{\rm c})$ diagram, however the lack of a population of
individual Li-rich objects towards the faint end of our sample places a
very strong {\em lower limit} to the cluster age of 35\,Myr. The detection
of Li in the {\em averaged} spectra of our faintest objects, together
with the expected low-levels of contamination by non-members, lead us
to believe that the LDB actually lies at $17.8 \leq I_{\rm c} < 18.3$ and
hence an upper limit to the LDB age of 54\,Myr.

The age of NGC\,2547 judged from isochrones in the $V/(V-I_{\rm c})$ and
$I_{\rm c}/(R-I_{\rm c})$ colour-magnitude diagrams is between 20 and
35\,Myr using the same evolutionary models from which the LDB ages were
determined. The sense and size of this discrepancy between the LDB and
low-mass isochronal ages is similar to that seen in another young
cluster, IC\,2391 and in the low-mass PMS binary system GJ\,871.1AB.
We suggest that inclusion of rotation or internal magnetic fields in
PMS evolutionary models {\em may} be able to reconcile the two age
determinations, at the expense of increasing the ages to values even
higher than the LDB ages presently indicate. Alternatively, some
mechanism must be identified which can increase the rate of Li depletion among
young, very low-mass fully convective PMS stars.

\begin{table*}
\begin{minipage}{135mm}
\caption{Objects with no detected Li feature. Columns\,1$-$10 as in Table\,\ref{table_li}. Column\,11 gives the estimated 2$-\sigma$ upper limits.} 
\label{table}
\begin{tabular}{rccllllllcc}
\hline
       & \multicolumn{2}{c}{J2000 Position} & \multicolumn{2}{c}{I$_{\rm c}$-band} & \multicolumn{2}{c}{$(R-I_{\rm c}$)} & \multicolumn{2}{c}{$(I_{\rm c}-Z$)}  &  & \multicolumn{1}{c}{Li}\\
        & ra & dec & mag & error & mag & error & mag & error & flag & \multicolumn{1}{c}{$2\sigma$ (\AA)}\\
\hline
  1 & 8 09 49.46 & $-$49 16 19.0 & 15.681 & 0.004 & 1.762 & 0.005 & 0.467 & 0.005& M& 0.14\\
  2 & 8 10 48.36 & $-$49 15 25.1 & 15.712 & 0.003 & 1.490 & 0.004 & 0.377 & 0.003& M& 0.09\\
  3 & 8 09 13.92 & $-$49 13 18.8 & 15.736 & 0.008 & 1.429 & 0.016 & 0.400 & 0.009& M& 0.10\\
  4 & 8 10 32.59 & $-$49 09 43.8 & 15.760 & 0.004 & 1.366 & 0.005 & 0.345 & 0.005& M& 0.08\\
  5 & 8 10 57.20 & $-$49 23 47.9 & 15.804 & 0.003 & 1.702 & 0.004 & 0.461 & 0.004& M& 0.09\\
  6 & 8 10 59.43 & $-$49 08 33.1 & 15.830 & 0.007 & 1.474 & 0.013 & 0.418 & 0.008& M& 0.12\\
  7 & 8 09 54.29 & $-$49 08 41.9 & 15.841 & 0.008 & 1.484 & 0.008 & 0.379 & 0.009& M& 0.13\\
  8 & 8 10 01.44 & $-$48 56 21.9 & 15.911 & 0.004 & 1.283 & 0.005 & 0.394 & 0.004& M& 0.11\\
  9 & 8 12 56.77 & $-$49 11 22.5 & 15.922 & 0.005 & 1.310 & 0.005 & 0.341 & 0.005& M& 0.11\\
 10 & 8 10 35.82 & $-$49 20 05.6 & 15.922 & 0.003 & 1.495 & 0.005 & 0.371 & 0.004& M& 0.08\\
 11 & 8 10 27.41 & $-$49 20 19.9 & 15.948 & 0.003 & 1.525 & 0.005 & 0.386 & 0.004& M& 0.10\\
 12 & 8 09 00.42 & $-$49 12 59.6 & 16.002 & 0.009 & 1.556 & 0.021 & 0.454 & 0.010& M& 0.10\\
 13 & 8 08 53.44 & $-$49 13 26.8 & 16.029 & 0.009 & 1.564 & 0.021 & 0.444 & 0.011& M& 0.15\\
 14 & 8 10 07.08 & $-$49 12 59.9 & 16.032 & 0.005 & 1.529 & 0.006 & 0.410 & 0.006& M& 0.10\\
 15 & 8 10 18.40 & $-$49 26 31.3 & 16.042 & 0.003 & 1.709 & 0.005 & 0.472 & 0.004& M& 0.12\\
 16 & 8 10 13.80 & $-$49 18 37.0 & 16.050 & 0.003 & 1.539 & 0.004 & 0.397 & 0.004& M& 0.15\\
 17 & 8 12 56.81 & $-$49 01 37.3 & 16.079 & 0.003 & 1.463 & 0.004 & 0.405 & 0.004& M& 0.11\\
 18 & 8 10 42.49 & $-$49 24 55.4 & 16.149 & 0.003 & 1.714 & 0.005 & 0.467 & 0.003& M& 0.12\\
 19 & 8 09 21.90 & $-$49 11 35.2 & 16.170 & 0.010 & 1.639 & 0.025 & 0.460 & 0.012& M& 0.10\\
 20 & 8 10 50.13 & $-$49 22 47.7 & 16.196 & 0.003 & 1.618 & 0.004 & 0.449 & 0.003& M& 0.16\\
 21 & 8 10 06.17 & $-$49 25 59.0 & 16.199 & 0.004 & 1.700 & 0.006 & 0.452 & 0.004& M& 0.14\\
 22 & 8 11 16.67 & $-$49 22 29.2 & 16.245 & 0.004 & 1.595 & 0.005 & 0.419 & 0.005& M& 0.64\\
 23 & 8 12 03.84 & $-$49 12 04.7 & 16.248 & 0.006 & 1.627 & 0.007 & 0.426 & 0.006& M& 0.13\\
 24 & 8 08 58.56 & $-$49 11 17.3 & 16.313 & 0.011 & 1.599 & 0.027 & 0.434 & 0.013& M& 0.14\\
 25 & 8 11 09.33 & $-$49 11 53.6 & 16.316 & 0.006 & 1.641 & 0.007 & 0.450 & 0.006& M& 0.20\\
 27 & 8 08 54.64 & $-$49 27 03.5 & 16.329 & 0.004 & 1.635 & 0.006 & 0.446 & 0.005& M& 0.14\\
 28 & 8 12 11.63 & $-$49 12 36.2 & 16.332 & 0.006 & 1.720 & 0.010 & 0.504 & 0.009& M& 0.21\\
 29 & 8 10 36.75 & $-$48 58 39.1 & 16.345 & 0.003 & 1.659 & 0.004 & 0.443 & 0.004& M& 0.40\\
 30 & 8 10 00.84 & $-$49 09 10.0 & 16.349 & 0.006 & 1.756 & 0.008 & 0.497 & 0.007& M& 0.15\\
 31 & 8 08 11.66 & $-$49 09 16.6 & 16.352 & 0.003 & 1.637 & 0.005 & 0.419 & 0.004& M& 0.19\\
 32 & 8 09 06.82 & $-$49 15 34.9 & 16.367 & 0.007 & 1.662 & 0.010 & 0.478 & 0.007& M& 0.15\\
 33 & 8 09 25.74 & $-$49 03 15.7 & 16.374 & 0.010 & 1.590 & 0.017 & 0.447 & 0.011& M& 0.14\\
 34 & 8 09 37.17 & $-$49 05 59.6 & 16.397 & 0.005 & 1.562 & 0.008 & 0.411 & 0.006& M& 0.12\\
 35 & 8 09 11.71 & $-$48 57 38.6 & 16.439 & 0.005 & 1.568 & 0.007 & 0.407 & 0.004& M& 0.12\\
 36 & 8 07 47.51 & $-$49 11 09.5 & 16.448 & 0.004 & 1.805 & 0.007 & 0.519 & 0.005& M& 0.33\\
 37 & 8 11 14.31 & $-$49 10 24.6 & 16.475 & 0.006 & 1.436 & 0.007 & 0.371 & 0.007& M& 0.12\\
 39 & 8 10 07.81 & $-$49 01 06.4 & 16.498 & 0.010 & 1.510 & 0.015 & 0.371 & 0.011& M& 0.17\\
 40 & 8 11 31.92 & $-$49 11 53.0 & 16.500 & 0.011 & 1.712 & 0.024 & 0.518 & 0.011& M& 0.15\\
 41 & 8 09 17.72 & $-$49 08 34.6 & 16.525 & 0.004 & 1.751 & 0.006 & 0.482 & 0.005& M& 0.17\\
 42 & 8 10 50.49 & $-$49 16 24.0 & 16.538 & 0.005 & 1.604 & 0.006 & 0.441 & 0.005& M& 0.23\\
 43 & 8 09 33.92 & $-$49 38 45.4 & 16.562 & 0.004 & 1.797 & 0.008 & 0.483 & 0.004& M& 0.24\\
 44 & 8 09 29.04 & $-$49 14 33.0 & 16.627 & 0.013 & 1.648 & 0.027 & 0.465 & 0.015& M& 0.13\\
 45 & 8 10 21.87 & $-$49 00 07.0 & 16.666 & 0.006 & 1.649 & 0.009 & 0.438 & 0.007& M& 0.22\\
 46 & 8 10 11.76 & $-$49 21 01.9 & 16.670 & 0.011 & 1.918 & 0.021 & 0.615 & 0.011& M& 0.23\\
 47 & 8 08 51.49 & $-$49 12 58.9 & 16.680 & 0.015 & 1.486 & 0.035 & 0.425 & 0.017& M& 0.15\\
 48 & 8 09 12.08 & $-$49 10 15.4 & 16.747 & 0.015 & 1.531 & 0.037 & 0.453 & 0.017& M& 0.12\\
 49 & 8 10 10.41 & $-$48 58 05.4 & 16.748 & 0.004 & 1.687 & 0.006 & 0.465 & 0.005& M& 0.20\\
 50 & 8 09 35.49 & $-$49 13 03.3 & 16.748 & 0.010 & 1.732 & 0.016 & 0.541 & 0.015& M& 0.16\\
 51 & 8 09 01.74 & $-$49 01 10.7 & 16.788 & 0.013 & 1.623 & 0.023 & 0.456 & 0.014& M& 0.39\\
 52 & 8 10 12.13 & $-$49 04 31.7 & 16.862 & 0.007 & 1.699 & 0.011 & 0.456 & 0.008& M& 0.22\\
 53 & 8 10 59.18 & $-$49 04 27.9 & 16.899 & 0.005 & 1.833 & 0.008 & 0.536 & 0.006& M& 0.35\\
 54 & 8 10 25.35 & $-$49 11 18.6 & 16.911 & 0.006 & 1.858 & 0.009 & 0.542 & 0.007& NM& 0.52\\
 55 & 8 10 23.69 & $-$48 59 35.5 & 16.925 & 0.007 & 1.659 & 0.011 & 0.465 & 0.008& M& 0.23\\
 56 & 8 10 02.44 & $-$49 05 13.3 & 16.938 & 0.007 & 1.640 & 0.011 & 0.442 & 0.008& M& 0.19\\
 57 & 8 11 31.07 & $-$49 14 10.6 & 16.948 & 0.009 & 1.618 & 0.011 & 0.447 & 0.009& NM& 0.35\\
 58 & 8 09 11.92 & $-$49 14 43.0 & 16.965 & 0.010 & 1.727 & 0.016 & 0.465 & 0.011& NM& 0.24\\
 59 & 8 11 39.94 & $-$49 03 25.9 & 16.977 & 0.005 & 1.707 & 0.008 & 0.489 & 0.007& M& 0.24\\
 60 & 8 10 49.78 & $-$49 08 19.9 & 16.977 & 0.004 & 1.699 & 0.006 & 0.476 & 0.005& M& 0.23\\
\hline
\end{tabular}
\end{minipage}
\end{table*}
\begin{table*}
\begin{minipage}{135mm}
\contcaption{Objects with no detected Li feature. Columns\,1$-$10 as in Table\,\ref{table_li}. Column\,11 gives the estimated 2$-\sigma$ upper limits.} 
\begin{tabular}{rccllllllcc}
\hline
       & \multicolumn{2}{c}{J2000 Position} & \multicolumn{2}{c}{I$_{\rm c}$-band} & \multicolumn{2}{c}{$(R-I_{\rm c}$)} & \multicolumn{2}{c}{$(I_{\rm c}-Z$)} &  & \multicolumn{1}{c}{Li}\\
        & ra & dec & mag & error & mag & error & mag & error & flag & \multicolumn{1}{c}{$2\sigma$ (\AA)}\\
 \hline
 61 & 8 09 09.26 & $-$49 20 42.7 & 16.989 & 0.006 & 1.638 & 0.008 & 0.470 & 0.007& M& 0.38\\
 62 & 8 08 38.74 & $-$49 12 49.1 & 17.018 & 0.005 & 1.811 & 0.009 & 0.477 & 0.007& M& 0.24\\
 63 & 8 08 20.49 & $-$49 04 54.7 & 17.069 & 0.007 & 1.730 & 0.011 & 0.450 & 0.008& M& 0.52\\
 64 & 8 08 44.32 & $-$49 20 12.5 & 17.102 & 0.007 & 1.699 & 0.010 & 0.491 & 0.008& M& 0.23\\
 66 & 8 10 17.67 & $-$49 23 29.7 & 17.157 & 0.006 & 1.624 & 0.011 & 0.429 & 0.008& M& 0.24\\
 68 & 8 10 46.82 & $-$49 03 47.8 & 17.179 & 0.005 & 1.743 & 0.008 & 0.497 & 0.006& M& 0.31\\
 69 & 8 11 16.74 & $-$49 06 56.7 & 17.183 & 0.005 & 1.683 & 0.007 & 0.480 & 0.006& M& 0.23\\
 70 & 8 07 41.00 & $-$49 19 07.8 & 17.187 & 0.005 & 1.600 & 0.007 & 0.414 & 0.006& M& 1.10\\
 71 & 8 10 04.60 & $-$49 11 39.0 & 17.192 & 0.011 & 1.753 & 0.015 & 0.494 & 0.013& M& 0.23\\
 73 & 8 10 46.88 & $-$49 21 34.1 & 17.211 & 0.005 & 1.876 & 0.009 & 0.586 & 0.006& M& 0.53\\
 74 & 8 12 04.66 & $-$49 02 17.6 & 17.219 & 0.006 & 1.704 & 0.010 & 0.513 & 0.008& M& 0.40\\
 76 & 8 10 12.75 & $-$49 01 49.8 & 17.255 & 0.017 & 1.825 & 0.030 & 0.566 & 0.017& M& 0.23\\
 77 & 8 10 23.18 & $-$49 04 04.2 & 17.265 & 0.017 & 1.763 & 0.031 & 0.472 & 0.018& NM& 0.68\\
 78 & 8 10 43.13 & $-$48 56 47.7 & 17.360 & 0.004 & 1.943 & 0.008 & 0.623 & 0.005& NM& 1.48\\
 79 & 8 08 17.93 & $-$49 04 17.7 & 17.405 & 0.018 & 1.728 & 0.030 & 0.489 & 0.019& M& 0.44\\
 80 & 8 10 46.81 & $-$49 27 45.5 & 17.412 & 0.004 & 1.769 & 0.007 & 0.488 & 0.005& M& 0.38\\
 81 & 8 11 03.97 & $-$49 10 01.9 & 17.472 & 0.012 & 1.712 & 0.016 & 0.485 & 0.013& NM& 0.20\\
 82 & 8 08 54.09 & $-$49 21 04.8 & 17.509 & 0.009 & 1.754 & 0.013 & 0.516 & 0.010& M& 0.33\\
 83 & 8 09 53.33 & $-$48 55 37.7 & 17.511 & 0.010 & 1.796 & 0.013 & 0.508 & 0.011& M& 0.36\\
 84 & 8 11 32.41 & $-$48 58 51.2 & 17.527 & 0.006 & 1.779 & 0.009 & 0.498 & 0.006& M& 0.34\\
 85 & 8 10 12.07 & $-$49 17 41.4 & 17.541 & 0.008 & 1.719 & 0.012 & 0.526 & 0.009& NM& 0.22\\
 86 & 8 07 29.69 & $-$49 02 03.0 & 17.599 & 0.010 & 1.809 & 0.018 & 0.496 & 0.011& M& 0.45\\
 87 & 8 09 12.40 & $-$49 22 28.0 & 17.639 & 0.009 & 1.675 & 0.014 & 0.438 & 0.011& M& 0.28\\
 88 & 8 09 33.84 & $-$49 30 32.6 & 17.658 & 0.007 & 1.949 & 0.012 & 0.607 & 0.008& M& 0.40\\
 89 & 8 10 16.48 & $-$49 03 36.6 & 17.672 & 0.012 & 1.866 & 0.022 & 0.564 & 0.014& M& 0.37\\
 90 & 8 08 58.32 & $-$49 30 09.2 & 17.710 & 0.007 & 1.987 & 0.012 & 0.594 & 0.008& M& 0.64\\
 92 & 8 10 57.64 & $-$49 19 58.1 & 17.808 & 0.012 & 1.808 & 0.022 & 0.510 & 0.015& NM& 0.30\\
 93 & 8 08 36.32 & $-$49 19 06.3 & 17.822 & 0.009 & 1.705 & 0.016 & 0.496 & 0.010& NM& 0.43\\
 94 & 8 10 41.23 & $-$49 12 22.1 & 17.850 & 0.016 & 1.761 & 0.023 & 0.567 & 0.016& NM& 0.21\\
 95 & 8 12 10.01 & $-$49 05 20.9 & 17.917 & 0.010 & 1.964 & 0.019 & 0.608 & 0.012& M& 1.30\\
 96 & 8 11 25.29 & $-$49 26 35.4 & 17.925 & 0.012 & 1.844 & 0.019 & 0.487 & 0.014& M& 0.35\\
 97 & 8 09 44.64 & $-$49 28 57.6 & 17.980 & 0.006 & 1.900 & 0.013 & 0.543 & 0.008& M& 0.68\\
 98 & 8 10 04.21 & $-$49 00 29.0 & 18.014 & 0.015 & 1.834 & 0.029 & 0.503 & 0.017& M& 0.35\\
 99 & 8 10 31.84 & $-$48 58 08.8 & 18.053 & 0.016 & 1.718 & 0.027 & 0.548 & 0.019& M& 0.46\\
101 & 8 10 53.27 & $-$49 16 48.6 & 18.095 & 0.011 & 1.914 & 0.018 & 0.604 & 0.012& NM& 0.76\\
102 & 8 09 42.58 & $-$49 04 39.6 & 18.143 & 0.012 & 1.736 & 0.022 & 0.482 & 0.013& NM& 0.30\\
103 & 8 12 00.17 & $-$49 18 57.6 & 18.163 & 0.011 & 1.901 & 0.019 & 0.536 & 0.013& M& 0.96\\
104 & 8 11 58.97 & $-$49 01 59.8 & 18.172 & 0.009 & 1.774 & 0.015 & 0.504 & 0.011& M& 0.61\\
105 & 8 12 59.18 & $-$49 04 02.0 & 18.224 & 0.013 & 1.782 & 0.022 & 0.522 & 0.016& NM& 0.88\\
106 & 8 09 08.08 & $-$49 21 59.6 & 18.233 & 0.014 & 1.885 & 0.023 & 0.590 & 0.016& NM& 0.68\\
107 & 8 11 01.18 & $-$48 59 08.4 & 18.261 & 0.016 & 1.994 & 0.045 & 0.589 & 0.017& M& 0.67\\
108 & 8 08 06.81 & $-$49 00 10.2 & 18.285 & 0.017 & 1.871 & 0.031 & 0.525 & 0.018& M& 0.85\\
109 & 8 08 03.61 & $-$49 05 06.6 & 18.385 & 0.018 & 1.981 & 0.037 & 0.579 & 0.020& NM& 1.19\\
\hline
\end{tabular}
\end{minipage}
\end{table*}

\section*{Acknowledgements}
We would like to thank: the director and staff of the Cerro Tololo
Interamerican Observatory, operated by the Association of Universities
for Research in Astronomy, Inc., under contract to the US National
Science Foundation; the director and staff of the Anglo Australian
Observatory and particularly Terry Bridges who assisted with the 2dF
observations; Nigel Hambly for providing accurate astrometric positions
for 2dF targets from the SuperCOSMOS sky survey. Computational work
was performed on the Keele node of the PPARC funded Starlink
network. JMO acknowledges support of the UK Particle Physics and
Astronomy Research Council and CRD acknowledges the support of the 
Nuffield Foundation in the form of an undergraduate research bursary.

\bsp

\label{lastpage}


\begin{thebibliography}{99}

\bibitem[\protect\citeauthoryear{Allen}{1996}]{allen96}
Allen L.E., 1996, Ph.D. thesis, Univ. Massachusetts

\bibitem[\protect\citeauthoryear{Allen \& Strom}{1995}]{allen95}
Allen L.E., Strom K.M., 1995, AJ, 109, 1379

\bibitem[\protect\citeauthoryear{Baraffe et al.}{1998}]{baraffe98}
Baraffe I., Chabrier G., Allard F., Hauschildt P.H., 1998, A\&A, 337, 403

\bibitem[\protect\citeauthoryear{Baraffe et al.}{2002}]{baraffe02}
Baraffe I., Chabrier G., Allard F., Hauschildt P.H., 2002, A\&A,
382, 563

\bibitem[\protect\citeauthoryear{Barrado y Navascu\'{e}s \&
Stauffer}{2003}]{navascues03}
Barrado y Navascu\'{e}s D., Stauffer J.R., 2003, IAU Symp.\,211 on ``Brown Dwarfs'', in ASP Conference Series, ed. E. Mart\'{\i}n, in press

\bibitem[\protect\citeauthoryear{Barrado y Navascu\'{e}s, Stauffer \&
Patten}{Barrado y Navascu\'{e}s et al.}{1999}]{navascues99} 
Barrado y Navascu\'{e}s D., Stauffer J.R., Patten B.M., 1999, ApJ, 522, 53

\bibitem[\protect\citeauthoryear{Barrado y Navascu\'{e}s et
al.}{2001}]{navascues01} 
Barrado y Navascu\'{e}s D., Stauffer J.R., Brice\~{n}o C. et al., 1999, ApJ,
522, 53

\bibitem[\protect\citeauthoryear{Bessell}{1979}]{bessell79}
Bessell M.S., 1979, PASP, 91, 589

\bibitem[\protect\citeauthoryear{Bildsten et al.}{1997}]{bildsten97} 
Bildsten L., Brown E.F., Matzner C.D., Ushomirsky G., 1997, ApJ, 482, 442

\bibitem[\protect\citeauthoryear{Brice\~{n}o et al.}{1998}]{briceno98}
Brice\~{n}o C., Hartmann L., Stauffer J., Mart\'{\i}n E., 1998, AJ, 115, 2074

\bibitem[\protect\citeauthoryear{Burke \& Pinsonneault}{2000}]{burke00}
Burke C.J., Pinsonneault M.H., 2000, AAS abstract, 197, 41.10

\bibitem[\protect\citeauthoryear{Chabrier \& Baraffe}{1997}]{chabrier97}
Chabrier G., Baraffe I., 1997, A\&A, 327, 1039

\bibitem[\protect\citeauthoryear{Clari\'{a}}{1982}]{claria82} 
Clari\'{a} J.J., 1982, A\&AS, 47, 323

\bibitem[\protect\citeauthoryear{D'Antona et al.}{2000}]{dantona00}
D'Antona F., Ventura P., Mazzitelli I., 2000, ApJ, L77

\bibitem[\protect\citeauthoryear{D'Antona \& Mazzitelli}{1997}]{dantona97}
D'Antona F., Mazzitelli I., 1997, Mem. Soc. Astr. It., 68, 807

\bibitem[\protect\citeauthoryear{Hambly et al.}{2001a}]{hambly01a}
Hambly N.C., MacGillivray H.T., Read M.A. et al., 2001a, MNRAS, 326, 1279

\bibitem[\protect\citeauthoryear{Hambly et al.}{2001b}]{hambly01b}
Hambly N.C., Davenhall A.C., Irwin M.J., MacGillivray H.T., 2001b, MNRAS, 326,
1315

\bibitem[\protect\citeauthoryear{Jeffries \& Tolley}{1998}]{jeffries98} 
Jeffries R.D., Tolley A.J., 1998, MNRAS, 300, 331

\bibitem[\protect\citeauthoryear{Jeffries, Totten \& James}{Jeffries et
al.}{2000a}]{jeffries00a} 
Jeffries R.D., Totten E.J., James D.J., 2000a, MNRAS, 316, 950 

\bibitem[\protect\citeauthoryear{Jeffries et al.}{2000b}]{jeffries00b} 
Jeffries R.D., Totten E.J., Barrado y Navascu\'{e}s D., Stauffer J.R., Hambly
N.C., 2000b, in ``Stellar clusters and associations: Convection, rotation and
dynamos'', ASP Conference Series, Vol. 198, eds. Pallavicini R., Micela G., San
Francisco, p. 281

\bibitem[\protect\citeauthoryear{Jeffries, Thurston \& Hambly}{Jeffries et al.}{2001a}]{jeffries01a} 
Jeffries R.D., Thurston M., Hambly N., 2001a, A\&A, 375, 863

\bibitem[\protect\citeauthoryear{Jeffries \& Naylor}{2001b}]{jeffries01b} 
Jeffries R.D., Naylor T., 2001b, in ``From Darkness to Light: Origin and
Evolution of Young Stellar Clusters'' ASP Conference Proceedings, Vol. 243, eds. T. Montmerle, Ph. Andre, p.633

\bibitem[\protect\citeauthoryear{Jeffries et al.}{2003}]{jeffries02} 
Jeffries R.D., Oliveira J.M., Barrado y Navascu\'{e}s D., Stauffer J.R., 2003, MNRAS, submitted

\bibitem[\protect\citeauthoryear{Landolt}{1992}]{landolt92}
Landolt A.U., 1992, AJ, 104, 340

\bibitem[\protect\citeauthoryear{Leggett}{1992}]{leggett92}
Leggett S.K., 1992, ApJS, 82, 351

\bibitem[\protect\citeauthoryear{Leggett et al.}{1996}]{leggett96}
Leggett S.K., Allard F., Berriman G. et al., 1996, ApJS, 104, 117

\bibitem[\protect\citeauthoryear{Lewis et al.}{2002}]{lewis02}
Lewis I.J., Cannon R.D., Taylor K. et al., 2002, MNRAS, 333, 279

\bibitem[\protect\citeauthoryear{Littlefair et al.}{2003}]{littlefair03} 
Littlefair S.P., Naylor T., Jeffries R.D., Devey C.R., 2003, MNRAS
submitted

\bibitem[\protect\citeauthoryear{Mazzei \& Pigatto}{1988}]{mazzei88} 
Mazzei P., Pigatto L., 1988, A\&A, 193, 148

\bibitem[\protect\citeauthoryear{Mermilliod}{1981}]{mermilliod81} 
Mermilliod J.C.,1981, A\&A, 97, 235

\bibitem[\protect\citeauthoryear{Meynet, Mermilliod \& Maeder}{Meynet et
al.}{1993}]{meynet93} 
Meynet G., Mermilliod J.C., Maeder A., 1993, A\&AS, 98, 477

\bibitem[\protect\citeauthoryear{Monet}{1998}]{monet98}
Monet D., 1998, BAAS, 30, 1427

\bibitem[\protect\citeauthoryear{Montes et al.}{1997}]{montes97}
Montes D., Mart\'{\i}n E.L., Fern\'{a}ndez-Figueroa M.J., Cornide M., de Castro
E., 1997, A\&AS, 123, 473

\bibitem[\protect\citeauthoryear{Naylor et al.}{2002}]{naylor02}
Naylor T., Totten E.J., Jeffries R.D., Pozzo M., Devey C.R., Thompson S.A.,
2002, MNRAS, 335, 291

\bibitem[\protect\citeauthoryear{Oppenheimer et al.}{1997}]{oppenheimer97}
Oppenheimer B.R., Basri G., Nakajima T., Kulkarni S.R., 1997, AJ, 113, 296

\bibitem[\protect\citeauthoryear{Ortiz \& L\'{e}pine}{1993}]{ortiz93}
Ortiz R., L\'{e}pine J.R.D., 1993, A\&A, 279, 90

\bibitem[\protect\citeauthoryear{Patten \& Simon}{1996}]{patten96}
Patten B.M., Simon T., 1996, ApJS, 106, 489

\bibitem[\protect\citeauthoryear{Pavlenko \& Magazzu}{1995}]{pavlenko95}
Pavlenko Y.V., Magazzu A., 1996, A\&A, 311, 961

\bibitem[\protect\citeauthoryear{Pinsonneault et al.}{1998}]{pins98}
Pinsonneault M.H., Stauffer J.R., Soderblom D.R., King J.R.,
Hanson R.B., 1998, ApJ, 504, 170

\bibitem[\protect\citeauthoryear{Randich et al.}{2001}]{randich01}
Randich S., Pallavicini R., Meola G., Stauffer J.R., Balachandran S.C., 2001,
A\&A, 372, 862

\bibitem[\protect\citeauthoryear{Robichon et al.}{2000}]{robichon00}
Robichon N., Arenou F., Mermilliod J.-C., Turon C., 2000, A\&A,
345, 471

\bibitem[\protect\citeauthoryear{Schaller et al.}{1992}]{schaller92} 
Schaller G., Schaerer D., Meynet G., Maeder A., 1992, A\&AS, 96, 269

\bibitem[\protect\citeauthoryear{Siess, Dufour \& Forestini}{Siess et
al.}{2000}]{siess00} 
Siess L., Dufour E., Forestini M., 2000, A\&A, 358, 593

\bibitem[\protect\citeauthoryear{Simon \& Patten}{1998}]{simon98}
Simon T., Patten B.M., 1998, PASP, 110, 283

\bibitem[\protect\citeauthoryear{Song, Bessell \& Zuckerman}{Song et al.}{2002}]{song02}
Song I., Bessell M.S., Zuckerman B., 2002, ApJ, 581, L43

\bibitem[\protect\citeauthoryear{Stauffer et al.}{1997}]{stauffer97} 
Stauffer J.R., Hartmann L.W., Prosser C.F. et al., 1997, ApJ, 479, 776

\bibitem[\protect\citeauthoryear{Stauffer, Schultz \& Kirkpatrick}{Stauffer et
al.}{1998}]{stauffer98} 
Stauffer J.R., Schultz G., Kirkpatrick J.D., 1998, ApJ, 499, 199

\bibitem[\protect\citeauthoryear{Stauffer et al.}{1999}]{stauffer99} 
Stauffer J.R., Barrado y Navascu\'{e}s D., Bouvier J. et al., 1999, ApJ, 527, 219

%\bibitem[\protect\citeauthoryear{Stauffer et al.}{1989}]{stauffer89}
%Stauffer J., Hartmann L., Jones B.F., McNamara B.R., 1989, ApJ, 342, 285

\bibitem[\protect\citeauthoryear{Ushomirsky et al.}{1998}]{ushomirsky98}
Ushomirsky G., Matzner C.D., Brown E.F. et al., 1998, ApJ, 497, 253

\bibitem[\protect\citeauthoryear{Ventura et al.}{1998}]{ventura98}
Ventura P., Zeppieri A., Mazzitelli I., D'Antona F., 1998, A\&A, 334, 953 

\bibitem[\protect\citeauthoryear{Zapatero Osorio et al.}{1999}]{zapatero99}
Zapatero Osorio M.R., Rebolo R., Mart\'{\i}n E.L. et al., 1999, A\&AS, 134, 537

\bibitem[\protect\citeauthoryear{Zapatero Osorio et al.}{2002}]{zapatero02}
Zapatero Osorio M.R., B\'{e}jar V.J.S., Pavlenko Ya. et al., A\&A, 384 937


\end{thebibliography}
\end{document}